\documentclass[11pt,a4paper]{article}
\usepackage{jheppub}
\usepackage{bm}
\usepackage{microtype}
\usepackage{graphicx}
\graphicspath{{paper_figures/}{SectionV_Global_FFE_Output/}}
\newcommand{\order}{\mathcal O}
\newcommand{\dd}{\mathrm d}

\title{Model-dependent analytic spin–torsion corrections to Blandford–Znajek energy extraction in Einstein–Cartan gravity}

\author[a]{Jingxu Wu}
\emailAdd{wuxj@my.msu.ru}

\affiliation[a]{Faculty of Physics, Lomonosov Moscow State University,
Moscow 119991, Russia}

\author[b]{Liangyu Luo}
\affiliation[b]{International School, I.M. Sechenov First Moscow State Medical University,
Moscow 119048, Russia}

\author[c]{Zhenzhou Lei}

\author[c]{Xiao Heng}
\affiliation[c]{School of Mathematics and Physics, Xinjiang Institute of Engineering,
Urumqi 830023, China}

\abstract{
We investigate the leading near-horizon response of Blandford--Znajek energy
extraction to a compact, neutral spin-polarized source within minimally coupled
Einstein--Cartan--Dirac--Maxwell theory. Eliminating the algebraic contortion
yields an effective axial contact interaction, which we embed into a conserved,
anisotropic phenomenological completion. Working at leading order in the torsion
parameter $\epsilon_T$, spatial anisotropy $\xi$, and slow rotation $\chi=a/M$
on the fixed-ADM branch, we derive the modified energy extraction rate at
optimal load. We find that the leading-order power ratio $P_{\rm BZ}^{\rm EC}/P_{\rm BZ}^{\rm K}$
receives distinct contributions from rotational dragging ($\ell=1$) and
magnetostatic flux redistribution ($\ell=2$). In the isotropic limit ($\xi=0$),
the power is enhanced for a co-rotating completion and suppressed for a
counter-rotating one, whereas for $\xi\neq0$ the net shift depends on the polar
quadrupole response. We also formulate the generalized Znajek identity and
linearized Grad--Shafranov framework, demonstrating that undetermined load-factor
shifts leave the leading power coefficient invariant due to stationarity at the
matched load point.
}

\keywords{Black holes, classical theories of gravity, force-free electrodynamics, torsion gravity}

\begin{document}
\maketitle
\flushbottom

\section{Introduction}

The theoretical setting combines the Kerr geometry and black-hole
mechanics \cite{Kerr1963,BoyerLindquist1967,Carter1968,
BardeenPressTeukolsky1972,PenroseFloyd1971,Hawking1972,
BardeenCarterHawking1973,Hawking1975} with force-free plasma dynamics.
The Blandford--Znajek (BZ) mechanism extracts rotational energy from
a black hole through a stationary magnetically dominated
magnetosphere \cite{GoldreichJulian1969,Michel1973,
BlandfordZnajek1977}. Its power is controlled by the horizon angular
velocity, the magnetic flux threading the horizon, and the regularity
relation between the poloidal current and the horizon fields. The
last ingredient is the Znajek condition
\cite{Znajek1977,MacdonaldThorne1982,ThornePriceMacdonald1986}.
Recent kinetic and wave-propagation studies have independently
tested the Kerr energy flux and its force-free transport structure
\cite{MeringoloCamilloniRezzolla2025,KoideNodaTakahashi2025}.
Thus, even when the Maxwell field strength retains the minimal form
$F=\dd A$, a local change in the horizon geometry can alter the
extracted power.

Einstein--Cartan (EC) gravity provides a clean setting for such a
change. The tetrad and Lorentz connection are independent, and matter
spin sources torsion algebraically
\cite{Sciama1964,Kibble1961,Hehl1976,HehlDatta1971,Shapiro2002,
Hammond2002,Trautman2006}.  Eliminating minimally coupled Dirac
torsion gives a local four-fermion interaction; related fermionic,
fluid and cosmological realizations have been studied in
refs.~\cite{Gasperini1986,ObukhovKorotky1987,Poplawski2010,
FreidelMinicTakeuchi2005,Mercuri2006,PerezRovelli2006,
Kazmierczak2009,BojowaldDas2008,MagueijoZlosnikKibble2013}.
Recent work has also sharpened the Cauchy problem, covariant
Weyssenhoff-fluid dynamics and effective fluid descriptions of
self-gravitating Dirac fields
\cite{LuzMena2025,AgarwalCarloni2025,
VignoloDeMariaFabbriCarloni2025,AbboudGavassinoSinghSperanza2025}.
Torsion is therefore local and nonpropagating: it vanishes outside the
spin-polarized material, while the metric response generated by that
material can extend into the adjacent vacuum. This separation lets us
isolate a near-horizon engine effect without postulating long-range
torsion hair.

The force-free and numerical literature establishes the robustness
of BZ extraction across analytical, magnetohydrodynamic and kinetic
descriptions \cite{BlandfordPayne1982,Beskin2010,Punsly2001,
Komissarov2002,Komissarov2004,McKinneyGammie2004,McKinney2005,
KomissarovMcKinney2007,TchekhovskoyNarayanMcKinney2010,
TchekhovskoyNarayanMcKinney2011,McKinneyTchekhovskoyBlandford2012,
PennaNarayanSadowski2013}.  Covariant and exact-solution approaches
are developed in refs.~\cite{Uchida1997,GrallaJacobson2014,
BrennanGrallaJacobson2013,MenonDermer2005,MenonDermer2007,
TanabeNagataki2008,PanYu2015,LupsascaRodriguezStrominger2014,
GrallaLupsascaStrominger2016,TomaTakahara2014,
ContopoulosKazanasPapadopoulos2013,NathanailContopoulos2014}, while
recent global and kinetic studies clarify current sheets and pair
creation \cite{EastYang2018,ParfreyPhilippovCerutti2019,
CrinquandCeruttiPhilippovParfreyDubus2020,
MeringoloCamilloniRezzolla2025,KoideNodaTakahashi2025}.
Analytic BZ perturbations in a non-Kerr background continue to
illustrate why horizon and magnetospheric corrections must be kept
at a uniform order \cite{FengCaiYangZhang2026}.  Our purpose is narrower:
we isolate the analytic linear response generated by a localized EC
source instead of adding another global plasma simulation.

We construct the controlled part of that response through the chain
\begin{equation}
 \text{Cartan equation}\to U^{\rm spin}_{\mu\nu}
 \to h^{(T)}_{\mu\nu}\to\text{Znajek condition}
 \to\delta P_{\rm BZ}.
\end{equation}
The axial current is prescribed as a coarse-grained profile rather
than derived from a global Dirac spinor.  We therefore specify, as an
additional model input, every component of a conserved effective
stress tensor, including its rotational current.  This closes the
linearized Einstein system and makes the dependence on the source
completion explicit.  The torsion-producing matter is taken to be
electromagnetically neutral at the order evaluated; the current of the
BZ magnetosphere belongs to an independent force-free plasma sector.
A rigidly rotating spin-fluid interpretation
relates its amplitude to the matter angular velocity, but does not
remove its independent constitutive status
\cite{Hehl1976,ObukhovKorotky1987,AgarwalCarloni2025,
VignoloDeMariaFabbriCarloni2025}.  The calculation is first order in the torsion
amplitude and linear in the angular-anisotropy parameter.  It is
performed at leading slow-rotation order, with relative
$\order(\chi^2)$ corrections left in the controlled remainder.  The perturbation
construction follows the standard
black-hole framework
\cite{ReggeWheeler1957,Zerilli1970,Teukolsky1973,Moncrief1974,
Chandrasekhar1983,MartelPoisson2005,KodamaIshibashi2003,Poisson2004},
and the thermodynamic consistency check uses the covariant phase-space
and isolated-horizon results of
refs.~\cite{Wald1993,IyerWald1994,SudarskyWald1992,
AshtekarFairhurstKrishnan2000,WaldZoupas2000,HollandsWald2013}.
The main output is a conditional response functional
$C_{\rm BZ}[\tau_{\mu\nu}^{(T)}]$, evaluated for one explicitly
defined conserved source model, not a universal EC--Dirac prediction
and not a large numerical simulation.
Sections~\ref{sec:ECDM}--\ref{sec:near_horizon_geometry} derive the EC source and
its geometric response. Sections~\ref{sec:force_free} and
\ref{sec:BZ_power} derive the horizon regularity relation and the
leading power correction while displaying, but not claiming to solve,
the global force-free boundary problem.  Section~\ref{sec:analytic_limits}
establishes limits and sign criteria, and
section~\ref{sec:first_law} gives first-law bookkeeping and
electromagnetic consistency. Possible
applications to radio-galaxy engines are deferred to future dynamical
work.

\section{Einstein--Cartan--Dirac--Maxwell theory}
\label{sec:ECDM}

The conventions and algebraic elimination used in this section follow
the standard Einstein--Cartan and Hehl--Datta formulations
\cite{Sciama1964,Kibble1961,Hehl1976,HehlDatta1971,Shapiro2002}.

We work in the minimal Einstein--Cartan--Dirac--Maxwell framework,
taking the tetrad $e^{a}{}_{\mu}$ and the Lorentz connection
$\omega^{ab}{}_{\mu}=-\omega^{ba}{}_{\mu}$ as independent
gravitational variables. Throughout this work we use $c=\hbar=1$, metric signature $(-,+,+,+)$, and
\begin{equation}
    \kappa = 8\pi G .
\end{equation}
The spacetime metric is reconstructed from the tetrad according to
\begin{equation}
    g_{\mu\nu}
    =
    \eta_{ab}e^{a}{}_{\mu}e^{b}{}_{\nu},
    \qquad
    \eta_{ab}=\mathrm{diag}(-1,1,1,1),
\end{equation}
with
\begin{equation}
    e\equiv\det(e^{a}{}_{\mu})=\sqrt{-g}.
\end{equation}
The torsion and curvature two-forms are
\begin{align}
    T^{a}
    &=
    de^{a}
    +
    \omega^{a}{}_{b}\wedge e^{b},
    \label{eq:torsion_twoform}
    \\
    R^{ab}
    &=
    d\omega^{ab}
    +
    \omega^{a}{}_{c}\wedge\omega^{cb}.
    \label{eq:curvature_twoform}
\end{align}
Equivalently,
\begin{align}
    T^{a}{}_{\mu\nu}
    &=
    2\partial_{[\mu}e^{a}{}_{\nu]}
    +
    2\omega^{a}{}_{b[\mu}e^{b}{}_{\nu]},
    \\
    R^{ab}{}_{\mu\nu}
    &=
    2\partial_{[\mu}\omega^{ab}{}_{\nu]}
    +
    2\omega^{a}{}_{c[\mu}\omega^{cb}{}_{\nu]} .
\end{align}

Metric compatibility allows the Lorentz connection to be written as
\begin{equation}
    \omega^{ab}{}_{\mu}
    =
    \mathring{\omega}^{ab}{}_{\mu}
    +
    K^{ab}{}_{\mu},
    \label{eq:omega_split}
\end{equation}
where $\mathring{\omega}^{ab}{}_{\mu}$ is the torsion-free spin
connection and $K^{ab}{}_{\mu}$ is the contortion. The corresponding
affine connection is
\begin{equation}
    \Gamma^{\lambda}{}_{\mu\nu}
    =
    \mathring{\Gamma}^{\lambda}{}_{\mu\nu}
    +
    K^{\lambda}{}_{\mu\nu},
\end{equation}
so that
\begin{equation}
    T^{\lambda}{}_{\mu\nu}
    =
    K^{\lambda}{}_{\mu\nu}
    -
    K^{\lambda}{}_{\nu\mu}.
    \label{eq:T_K}
\end{equation}
The Riemann--Cartan curvature can then be decomposed as
\begin{align}
    R_{\alpha\beta\mu\nu}(\Gamma)
    =&\;
    \mathring{R}_{\alpha\beta\mu\nu}
    +
    \mathring{\nabla}_{\mu}K_{\alpha\beta\nu}
    -
    \mathring{\nabla}_{\nu}K_{\alpha\beta\mu}
    \nonumber\\
    &+
    K_{\alpha\rho\mu}K^{\rho}{}_{\beta\nu}
    -
    K_{\alpha\rho\nu}K^{\rho}{}_{\beta\mu}.
    \label{eq:R_K}
\end{align}

The microscopic action is chosen as
\begin{equation}
    S
    =
    \int d^{4}x\,e
    \left[
        \frac{1}{2\kappa}R(e,\omega)
        +
        \mathcal{L}_{D}
        -
        \frac14F_{\mu\nu}F^{\mu\nu}
    \right],
    \label{eq:ECDM_action}
\end{equation}
where the Hermitian Dirac Lagrangian is
\begin{equation}
    \mathcal{L}_{D}
    =
    \frac{i}{2}
    \left[
        \bar{\psi}\gamma^{\mu}D_{\mu}\psi
        -
        (D_{\mu}\bar{\psi})\gamma^{\mu}\psi
    \right]
    -
    m\bar{\psi}\psi .
    \label{eq:Dirac_L}
\end{equation}
We use
\begin{equation}
    \gamma^{\mu}=e_{a}{}^{\mu}\gamma^{a},
    \qquad
    \{
      \gamma^{a},
      \gamma^{b}
    \}
    =
    2\eta^{ab},
\end{equation}
and
\begin{equation}
    \gamma_{ab}
    =
    \frac12[\gamma_{a},\gamma_{b}].
\end{equation}
The spinor covariant derivatives are defined by
\begin{align}
    D_{\mu}\psi
    &=
    \left(
        \partial_{\mu}
        -
        \frac14\omega^{ab}{}_{\mu}\gamma_{ab}
        -
        iq_{\rm spin}A_{\mu}
    \right)\psi,
    \label{eq:Dpsi}
    \\
    D_{\mu}\bar{\psi}
    &=
    \partial_{\mu}\bar{\psi}
    +
    \frac14\omega^{ab}{}_{\mu}\bar{\psi}\gamma_{ab}
    +
    iq_{\rm spin}A_{\mu}\bar{\psi}.
\end{align}
Here $q_{\rm spin}$ denotes the electromagnetic charge of the
torsion-producing Dirac species; it is distinct from the radial
profile exponent $q$ introduced in section~\ref{sec:spin_source}.
The general action permits $q_{\rm spin}\ne0$, in which case variation
with respect to $A_\mu$ produces the vector current
$J_D^\mu=q_{\rm spin}\bar\psi\gamma^\mu\psi$.  To isolate the EC
geometric response studied in this paper, we select the neutral sector
\begin{equation}
 q_{\rm spin}=0,
 \qquad J_D^\mu=0.
 \label{eq:neutral_spin_source}
\end{equation}
The force-free current introduced in section~\ref{sec:force_free} is
therefore a separate plasma current, not a Dirac vector current of the
matter sourcing torsion.

The electromagnetic field strength is defined by the exterior
derivative,
\begin{equation}
    F=dA,
    \qquad
    F_{\mu\nu}
    =
    2\partial_{[\mu}A_{\nu]},
    \label{eq:F_dA}
\end{equation}
which preserves the standard local $U(1)$ symmetry. Consequently,
torsion does not enter $F_{\mu\nu}$ directly in the minimal model
considered here.

The independent variation with respect to the Lorentz connection is
algebraic. For a minimally coupled Dirac field, the spin density is
completely antisymmetric and excites only the axial component of the
torsion. Introducing
\begin{equation}
    J_{5}^{\mu}
    =
    \bar{\psi}\gamma^{\mu}\gamma^{5}\psi,
    \qquad
    \gamma^{5}
    =
    i\gamma^{0}\gamma^{1}\gamma^{2}\gamma^{3},
    \label{eq:J5}
\end{equation}
the contortion may be parameterized as
\begin{equation}
    K_{abc}
    =
    k_{A}\epsilon_{abcd}J_{5}^{d}.
    \label{eq:Kaxial}
\end{equation}
Its traces vanish identically,
\begin{equation}
    K^{a}{}_{ab}
    =
    K^{a}{}_{ba}
    =
    0.
\end{equation}

Using eq.~\eqref{eq:Kaxial} in the curvature decomposition yields
\begin{equation}
    R(e,\omega)
    =
    \mathring{R}
    +
    6k_{A}^{2}J_{5\mu}J_{5}^{\mu},
    \label{eq:R_axial}
\end{equation}
and hence
\begin{equation}
    \Delta\mathcal{L}_{\rm grav}
    =
    \frac{3}{\kappa}
    k_{A}^{2}
    J_{5\mu}J_{5}^{\mu}.
\end{equation}
The contortion-dependent contribution from the Dirac sector can be
reduced using
\begin{equation}
    \{
      \gamma^{c},
      \gamma^{ab}
    \}
    =
    2\gamma^{cab},
\end{equation}
with $\gamma^{abc}=\gamma^{[a}\gamma^{b}\gamma^{c]}$, giving
\begin{equation}
    \Delta\mathcal{L}_{D}
    =
    \frac32
    k_{A}
    J_{5\mu}J_{5}^{\mu}.
    \label{eq:LD_K}
\end{equation}
The complete algebraic dependence on the axial contortion is therefore
\begin{equation}
    \mathcal{L}_{K}
    =
    \frac{3}{\kappa}
    k_{A}^{2}J_{5}^{2}
    +
    \frac32k_{A}J_{5}^{2},
    \qquad
    J_{5}^{2}
    \equiv
    J_{5\mu}J_{5}^{\mu}.
    \label{eq:LK}
\end{equation}
The Cartan equation
\begin{equation}
    \frac{\partial\mathcal{L}_{K}}
         {\partial k_{A}}
    =
    0
\end{equation}
then gives
\begin{equation}
    k_{A}
    =
    -\frac{\kappa}{4},
\end{equation}
and consequently
\begin{equation}
   {
    K_{\alpha\beta\gamma}
    =
    -\frac{\kappa}{4}
    \epsilon_{\alpha\beta\gamma\delta}
    J_{5}^{\delta}
    }.
    \label{eq:Ksolution}
\end{equation}
This explicitly shows that the torsion is nondynamical and vanishes
whenever the local spin density vanishes.

Eliminating the connection gives the effective four-fermion
interaction
\begin{equation}
   {
    \mathcal{L}_{4\psi}
    =
    +\frac{3\kappa}{16}
    J_{5\mu}J_{5}^{\mu}
    }.
    \label{eq:L4psi}
\end{equation}
The original Riemann--Cartan system is therefore equivalent to the
torsion-free effective theory
\begin{align}
    S_{\rm eff}
    =
    \int d^{4}x\,e
    \Bigg[
        &
        \frac{1}{2\kappa}\mathring{R}
        +
        \mathring{\mathcal{L}}_{D}
        -
        \frac14F_{\mu\nu}F^{\mu\nu}
        \nonumber\\
        &
        +
        \frac{3\kappa}{16}
        J_{5\mu}J_{5}^{\mu}
    \Bigg].
    \label{eq:Seff}
\end{align}
Variation with respect to $\bar{\psi}$ gives the nonlinear
Hehl--Datta equation
\begin{equation}
    i\gamma^{\mu}\mathring{D}_{\mu}\psi
    -
    m\psi
    +
    \frac{3\kappa}{8}
    J_{5\mu}
    \gamma^{\mu}\gamma^{5}\psi
    =
    0.
    \label{eq:HD}
\end{equation}

The Maxwell energy-momentum tensor retains its standard form,
\begin{equation}
    T_{\mu\nu}^{\rm EM}
    =
    F_{\mu\rho}F_{\nu}{}^{\rho}
    -
    \frac14g_{\mu\nu}
    F_{\rho\sigma}F^{\rho\sigma},
    \label{eq:TEM}
\end{equation}
and satisfies
\begin{equation}
    g^{\mu\nu}T_{\mu\nu}^{\rm EM}
    =
    0.
\end{equation}
The torsion-free Dirac contribution is
\begin{align}
    T_{\mu\nu}^{D}
    =
    \frac{i}{4}
    \Big[
        &
        \bar{\psi}\gamma_{\mu}
        \mathring{D}_{\nu}\psi
        +
        \bar{\psi}\gamma_{\nu}
        \mathring{D}_{\mu}\psi
        \nonumber\\
        &
        -
        (\mathring{D}_{\mu}\bar{\psi})
        \gamma_{\nu}\psi
        -
        (\mathring{D}_{\nu}\bar{\psi})
        \gamma_{\mu}\psi
    \Big].
    \label{eq:TD}
\end{align}

For the effective decomposition adopted above, the explicit
four-fermion contribution to the tetrad equation can be written as
\begin{equation}
   {
    U_{\mu\nu}^{\rm spin}
    =
    +\frac{3\kappa}{16}
    g_{\mu\nu}
    J_{5\rho}J_{5}^{\rho}
    }.
    \label{eq:Uspin}
\end{equation}
Its trace is
\begin{equation}
    U^{\rm spin}
    =
    +\frac{3\kappa}{4}
    J_{5\rho}J_{5}^{\rho}.
\end{equation}
The effective Einstein equation therefore becomes
\begin{equation}
   {
    \mathring{G}_{\mu\nu}
    =
    \kappa
    \left(
        T_{\mu\nu}^{D}
        +
        T_{\mu\nu}^{\rm EM}
    \right)
    +
    \frac{3\kappa^{2}}{16}
    g_{\mu\nu}
    J_{5\rho}J_{5}^{\rho}
    }.
    \label{eq:Einstein_effective}
\end{equation}

\begin{figure}[t]
  \centering
  \includegraphics[width=.82\textwidth]{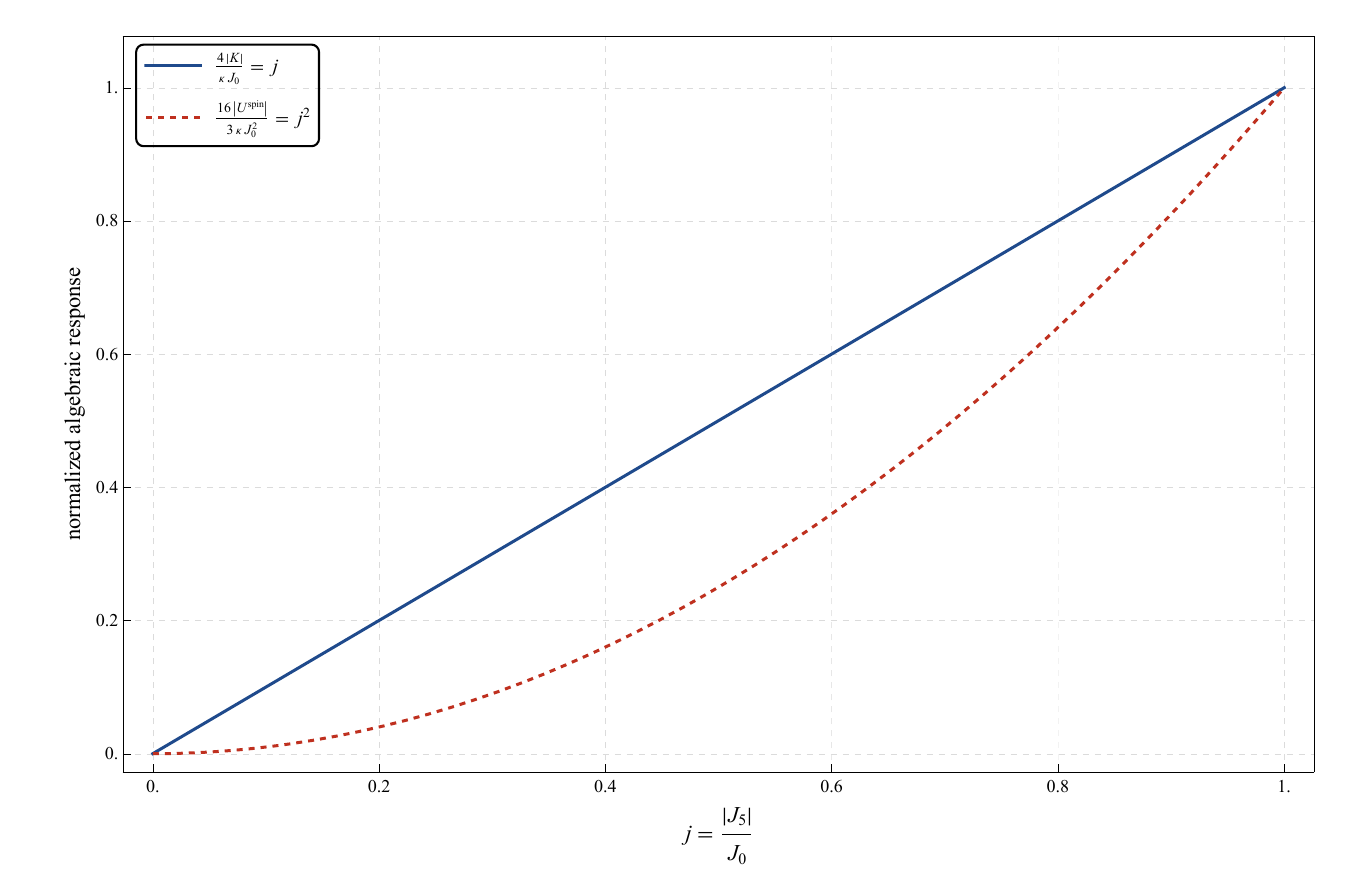}
  \caption{Normalized algebraic response to the axial-current
  amplitude $j=|J_5|/J_0$.  The contortion amplitude is linear in
  $j$, whereas the effective spin stress is quadratic.  Both vanish
  locally when the spin source is removed, illustrating the
  nonpropagating character of minimal Einstein--Cartan torsion.}
  \label{fig:II_algebraic_scaling}
\end{figure}

The different perturbative orders of the two algebraic responses are
displayed in figure~\ref{fig:II_algebraic_scaling}.  The figure is a
plot of the exact local scalings implied by the Cartan equation and
eq.~\eqref{eq:Uspin}; it is not a numerical solution of the field
equations.

Equation~\eqref{eq:Einstein_effective} is the form required in the
subsequent black-hole calculation. The torsion correction is local,
quadratic in the axial spin density, and enters the metric field
equation at order $\kappa^{2}J_{5}^{2}$. In particular,
\begin{equation}
    J_{5}^{\mu}=0
    \quad\Longrightarrow\quad
    K_{\alpha\beta\gamma}=0,
\end{equation}
so the minimal Einstein--Cartan theory does not introduce an
independent long-range torsion field in vacuum. The modification of
the Blandford--Znajek process considered below is therefore a
near-horizon source effect: spin-polarized matter changes the local
geometry, which in turn modifies the horizon quantities and
force-free boundary conditions relevant for electromagnetic energy
extraction.

\section{Stationary axisymmetric near-horizon spin source}
\label{sec:spin_source}

The axial-current construction below is a coarse-grained realization
of spin-polarized matter, while the optional fluid interpretation uses
the Weyssenhoff--Frenkel closure
\cite{Hehl1976,ObukhovKorotky1987,KatkarPhadatare2025,
AgarwalCarloni2025,KatkarPhadatare2026}; it is not a solved global Dirac
configuration.  It is also electromagnetically neutral as specified in
eq.~\eqref{eq:neutral_spin_source}; its axial current sources torsion,
but it supplies no independent Maxwell vector current.

We now construct a stationary and axisymmetric spin-polarized source
localized in the near-horizon region of a rotating black hole.  The
purpose of this construction is not to introduce an independent
torsion hair in vacuum.  Rather, it provides a controlled realization
of the local axial current entering the Einstein--Cartan relation
\begin{equation}
    K_{\alpha\beta\gamma}
    =
    -\frac{\kappa}{4}
    \epsilon_{\alpha\beta\gamma\delta}
    J_{5}^{\delta},
    \label{eq:III_K_J5}
\end{equation}
derived in section~\ref{sec:ECDM}. The resulting source will serve as
the matter-sector input for the perturbative black-hole geometry
constructed below.

At zeroth order in the spin--torsion coupling we take the geometry to
be Kerr.  In Boyer--Lindquist coordinates
$x^\mu=(t,r,\theta,\phi)$,
\begin{align}
    d\bar{s}^{2}
    =&
    -\left(
        1-\frac{2Mr}{\Sigma}
    \right)dt^{2}
    -
    \frac{4Mar\sin^{2}\theta}{\Sigma}
    dt\,d\phi
    +
    \frac{\Sigma}{\Delta}dr^{2}
    +
    \Sigma\,d\theta^{2}
    \nonumber\\
    &
    +
    \sin^{2}\theta
    \left(
        r^{2}+a^{2}
        +
        \frac{2Ma^{2}r\sin^{2}\theta}{\Sigma}
    \right)d\phi^{2},
    \label{eq:III_Kerr}
\end{align}
where
\begin{equation}
    \Sigma
    =
    r^{2}+a^{2}\cos^{2}\theta,
    \qquad
    \Delta
    =
    r^{2}-2Mr+a^{2}.
    \label{eq:III_SigmaDelta}
\end{equation}
The two Killing horizons are located at
\begin{equation}
    r_{\pm}
    =
    M
    \pm
    \sqrt{M^{2}-a^{2}},
    \label{eq:III_rpm}
\end{equation}
and the unperturbed horizon angular velocity is
\begin{equation}
    \bar{\Omega}_{H}
    =
    \frac{a}{r_{+}^{2}+a^{2}}.
    \label{eq:III_OmegaH}
\end{equation}
A bar will be used throughout this section for quantities evaluated
on the Kerr background.

We parameterize the coarse-grained axial current as
\begin{equation}
    J_{5}^{\mu}
    =
    {\cal A}(r,\theta)\,
    s^{\mu},
    \label{eq:III_J5_general}
\end{equation}
where $s^{\mu}$ is a unit spacelike polarization vector satisfying
\begin{equation}
    \bar g_{\mu\nu}s^{\mu}s^{\nu}
    =
    1,
    \qquad
    u_{\mu}s^{\mu}
    =
    0.
    \label{eq:III_s_constraints}
\end{equation}
Here $u^\mu$ denotes the local matter four-velocity wherever such a
fluid description is appropriate.  For a stationary circular
configuration outside the horizon one may write
\begin{equation}
    u^{\mu}
    =
    u^{t}
    \left(
        1,0,0,\Omega
    \right),
    \qquad
    u^{t}
    =
    \left[
        -
        \left(
            \bar g_{tt}
            +
            2\Omega\bar g_{t\phi}
            +
            \Omega^{2}\bar g_{\phi\phi}
        \right)
    \right]^{-1/2}.
    \label{eq:III_circular_velocity}
\end{equation}
The explicit choice of $s^\mu$ is not required for the scalar
spin--spin contribution derived below, because that contribution
depends only on $J_{5\mu}J_{5}^{\mu}$.  A convenient local
representative used for symbolic checks is given in
Appendix~\ref{app:spin_source_checks}.

The scalar amplitude is chosen in the separable form
\begin{equation}
    {\cal A}(r,\theta)
    =
    J_{0}
    \left(
        \frac{r}{M}
    \right)^{q}
    {\cal W}_{\rm NH}(r)
    f(\theta),
    \label{eq:III_A}
\end{equation}
where $J_{0}$ fixes the characteristic axial-current density,
$q$ controls its radial weighting, and ${\cal W}_{\rm NH}$ confines
the source to a finite neighborhood of the outer horizon.  No
assumption on the sign of $q$ is required by the formal construction;
$q<0$ corresponds to an additional outward decrease of the spin
density.

For the angular dependence we take the lowest nontrivial
equatorially symmetric deformation,
\begin{equation}
    f(\theta)
    =
    1
    +
    \xi P_{2}(\cos\theta),
    \qquad
    P_{2}(x)
    =
    \frac12(3x^{2}-1).
    \label{eq:III_angular}
\end{equation}
The profile satisfies
\begin{equation}
    f(\pi-\theta)=f(\theta),
\end{equation}
and is therefore compatible with reflection symmetry across the
equatorial plane.  Requiring $f(\theta)\geq0$ for all
$\theta\in[0,\pi]$ gives
\begin{equation}
    -1
    \leq
    \xi
    \leq
    2.
    \label{eq:III_xi_range}
\end{equation}
The interval in eq.~\eqref{eq:III_xi_range} is only the mathematical
nonnegativity range of the unexpanded profile. Every result in this
paper uses the stricter perturbative domain $|\xi|\ll1$ and is
truncated at $\order(\xi)$, so that only monopolar and quadrupolar
sectors occur in the scalar/even-parity source.  The rotational
odd-parity source is decomposed separately into $\ell=1,3$ sectors
through $\order(\chi\xi)$ below.

To obtain compact radial support without introducing a discontinuity
at the outer edge of the source, define
\begin{equation}
    x
    \equiv
    \frac{r-r_{+}}{\Delta r},
    \qquad
    \Delta r>0,
\end{equation}
and choose
\begin{equation}
    {\cal W}_{\rm NH}(r)
    =
    \begin{cases}
    \displaystyle
    \exp\left[
        1-\frac{1}{1-x^{2}}
    \right],
    &
    0\leq x<1,
    \\[2ex]
    0,
    &
    x\geq1.
    \end{cases}
    \label{eq:III_window}
\end{equation}
The normalization has been selected such that
\begin{equation}
    {\cal W}_{\rm NH}(r_{+})=1,
    \label{eq:III_window_horizon}
\end{equation}
whereas
\begin{equation}
    {\cal W}_{\rm NH}(r)
    =
    0,
    \qquad
    r\geq r_{+}+\Delta r.
    \label{eq:III_window_vacuum}
\end{equation}
All derivatives approach zero at
$r=r_{+}+\Delta r$ from inside the source region.  Consequently the
source is smoothly matched to a torsion-free exterior rather than
terminated by a sharp surface layer.

\begin{figure}[t]
  \centering
  \includegraphics[width=\textwidth]{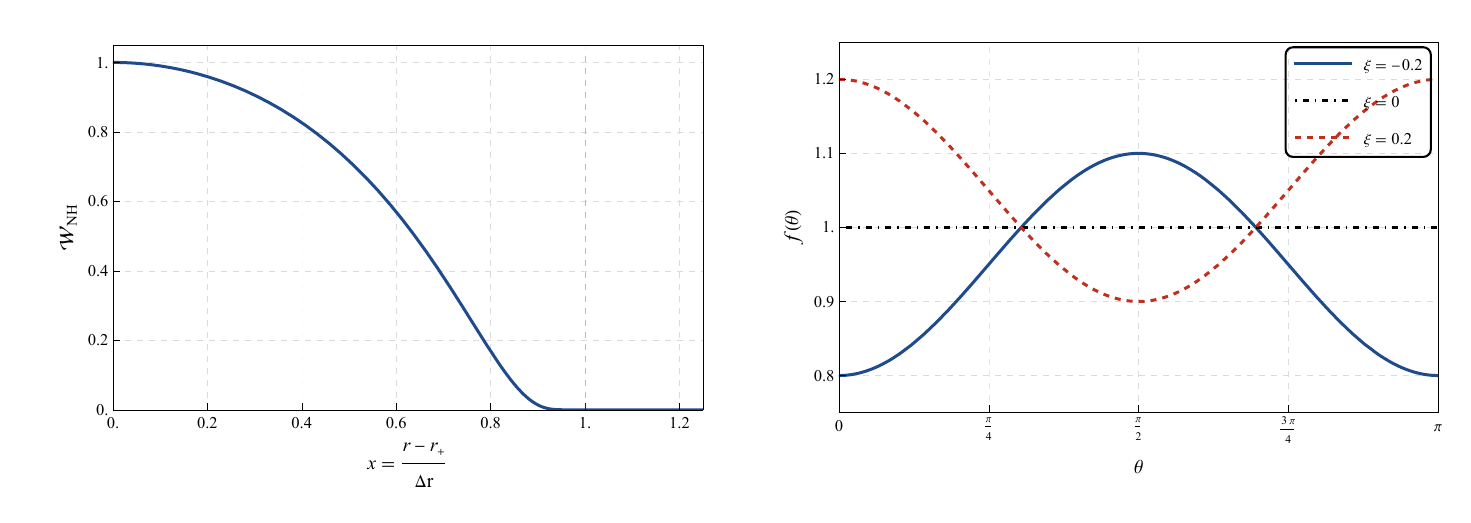}
  \caption{Analytic near-horizon source profile.  Left: the smooth
  compact-support window ${\cal W}_{\rm NH}$, normalized at the
  horizon and identically zero for $x\geq1$.  Right: the angular
  factor $f(\theta)=1+\xi P_2(\cos\theta)$ for
  $\xi=-0.2,0,0.2$. The curves are direct plots of
  eqs.~\eqref{eq:III_angular} and \eqref{eq:III_window}.}
  \label{fig:III_source_profile}
\end{figure}

Figure~\ref{fig:III_source_profile} collects the two independent
shape functions entering eq.~\eqref{eq:III_A}.  The plotted angular
values lie inside the actual perturbative domain.

Stationarity and axisymmetry are imposed at the level of the axial
current,
\begin{equation}
    {\cal L}_{\partial_t}J_{5}^{\mu}=0,
    \qquad
    {\cal L}_{\partial_\phi}J_{5}^{\mu}=0.
    \label{eq:III_symmetries}
\end{equation}
Using eq.~\eqref{eq:III_s_constraints}, its invariant norm becomes
particularly simple:
\begin{equation}
    J_{5}^{2}
    \equiv
    \bar g_{\mu\nu}
    J_{5}^{\mu}J_{5}^{\nu}
    =
    {\cal A}^{2}(r,\theta).
    \label{eq:III_J5sq_abstract}
\end{equation}
Hence
\begin{equation}
    {
    J_{5}^{2}
    =
    J_{0}^{2}
    \left(
        \frac{r}{M}
    \right)^{2q}
    {\cal W}_{\rm NH}^{2}(r)
    \left[
        1+\xi P_{2}(\cos\theta)
    \right]^{2}
    } .
    \label{eq:III_J5sq}
\end{equation}
In particular, its horizon value is finite,
\begin{equation}
    J_{5,H}^{2}
    =
    J_{0}^{2}
    \left(
        \frac{r_{+}}{M}
    \right)^{2q}
    \left[
        1+\xi P_{2}(\cos\theta)
    \right]^{2}.
    \label{eq:III_J5sq_horizon}
\end{equation}
The finiteness of this scalar, rather than the behavior of individual
Boyer--Lindquist components, is the relevant first test of horizon
regularity.  The corresponding tensorial quantities are regular when
expressed in horizon-penetrating coordinates, as discussed in
Appendix~\ref{app:spin_source_checks}.

Equation~\eqref{eq:III_K_J5} now gives the complete algebraic torsion
generated by the source,
\begin{equation}
    {
    K_{\alpha\beta\gamma}
    =
    -\frac{\kappa J_{0}}{4}
    \left(
        \frac{r}{M}
    \right)^{q}
    {\cal W}_{\rm NH}(r)
    f(\theta)
    \epsilon_{\alpha\beta\gamma\delta}
    s^{\delta}
    } .
    \label{eq:III_K_full}
\end{equation}
It follows immediately that
\begin{equation}
    r\geq r_{+}+\Delta r
    \quad\Longrightarrow\quad
    J_{5}^{\mu}=0
    \quad\Longrightarrow\quad
    K_{\alpha\beta\gamma}=0.
    \label{eq:III_vacuum_torsion}
\end{equation}
Thus the present model contains no torsion outside the material
source.  This statement concerns the torsion itself; the metric
perturbation generated by the source need not vanish outside its
support and will be determined by the gravitational field equations
in the following section.

The local four-fermion term derived in
section~\ref{sec:ECDM} produces
\begin{equation}
    U_{\mu\nu}^{\rm spin}
    =
    +\frac{3\kappa}{16}
    J_{5}^{2}
    \bar g_{\mu\nu}
    +
    {\cal O}(\epsilon_{T}^{2}),
    \label{eq:III_Uspin}
\end{equation}
where evaluating the metric on the Kerr background is sufficient at
first order in the spin--torsion perturbation.  It is convenient to
introduce
\begin{equation}
    {\cal S}(r,\theta)
    \equiv
    +\frac{3\kappa}{16}
    J_{5}^{2}(r,\theta),
    \label{eq:III_S}
\end{equation}
so that
\begin{equation}
    U_{\mu\nu}^{\rm spin}
    =
    {\cal S}\,
    \bar g_{\mu\nu},
    \qquad
    U^{\mu}{}_{\nu}{}^{\rm spin}
    =
    {\cal S}\,
    \delta^{\mu}{}_{\nu}.
    \label{eq:III_Umixed}
\end{equation}
Two useful scalar diagnostics are therefore
\begin{align}
    U^{\rm spin}
    &\equiv
    U^{\mu}{}_{\mu}{}^{\rm spin}
    =
    4{\cal S}
    =
    +\frac{3\kappa}{4}J_{5}^{2},
    \label{eq:III_Utrace}
    \\
    U_{\mu\nu}^{\rm spin}
    U_{\rm spin}^{\mu\nu}
    &=
    4{\cal S}^{2}
    =
    \frac{9\kappa^{2}}{64}
    \left(
        J_{5}^{2}
    \right)^{2}.
    \label{eq:III_Uquadratic}
\end{align}
For any unit timelike observer $u^\mu$ in the source region,
\begin{equation}
    \rho_{\rm spin}
    \equiv
    U_{\mu\nu}^{\rm spin}
    u^{\mu}u^{\nu}
    =
    -{\cal S}
    =
    -\frac{3\kappa}{16}J_{5}^{2}.
    \label{eq:III_rhospin}
\end{equation}
Since the polarization current used here is spacelike,
$J_{5}^{2}>0$, the isolated contact contribution has negative local
energy density in this convention. Only the conserved completion
$U_{\mu\nu}^{\rm spin}+\Theta_{\mu\nu}$ has an invariant thermodynamic
interpretation. On the
unperturbed horizon,
\begin{equation}
    \rho_{{\rm spin},H}
    =
    -\frac{3\kappa J_{0}^{2}}{16}
    \left(
        \frac{r_{+}}{M}
    \right)^{2q}
    \left[
        1+\xi P_{2}(\cos\theta)
    \right]^{2}.
    \label{eq:III_rhospin_horizon}
\end{equation}

The natural dimensionless strength of the source is
\begin{equation}
    {
    \epsilon_{T}
    \equiv
    \kappa^{2}J_{0}^{2}M^{2}
    } ,
    \label{eq:III_epsilonT}
\end{equation}
which compares the characteristic curvature generated by the
spin--spin term with the background Kerr curvature scale $M^{-2}$.
The corresponding local quantity is
\begin{align}
    \epsilon_{T}^{\rm loc}(r,\theta)
    &\equiv
    \kappa^{2}M^{2}J_{5}^{2}
    \nonumber\\
    &=
    \epsilon_{T}
    \left(
        \frac{r}{M}
    \right)^{2q}
    {\cal W}_{\rm NH}^{2}(r)
    \left[
        1+\xi P_{2}(\cos\theta)
    \right]^{2}.
    \label{eq:III_epsilon_local}
\end{align}
The perturbative regime is defined by
\begin{equation}
    \max_{\mathcal D_{\rm NH}}
    \epsilon_{T}^{\rm loc}
    \ll1,
    \label{eq:III_perturbative_condition}
\end{equation}
where
\begin{equation}
    {\cal D}_{\rm NH}
    =
    \left\{
       (r,\theta):
       r_{+}
       \leq r
       <
       r_{+}+\Delta r
    \right\}.
\end{equation}
In terms of $\epsilon_T$, the algebraic spin contribution to the
Einstein equation takes the compact form
\begin{equation}
    {
    \kappa U_{\mu\nu}^{\rm spin}
    =
    +
    \frac{3\epsilon_{T}}{16M^{2}}
    {\cal P}(r,\theta)
    \bar g_{\mu\nu}
    } ,
    \label{eq:III_Einstein_source}
\end{equation}
with
\begin{equation}
    {\cal P}(r,\theta)
    =
    \left(
        \frac{r}{M}
    \right)^{2q}
    {\cal W}_{\rm NH}^{2}(r)
    \left[
        1+\xi P_{2}(\cos\theta)
    \right]^{2}.
    \label{eq:III_Psource}
\end{equation}

The choice in eq.~\eqref{eq:III_angular} is particularly useful
because its multipolar content terminates at $\ell=4$:
\begin{align}
    \left[
        1+\xi P_{2}(\mu)
    \right]^{2}
    =&
    \left(
        1+\frac{\xi^{2}}{5}
    \right)P_{0}(\mu)
    \nonumber\\
    &
    +
    \left(
        2\xi+\frac{2\xi^{2}}{7}
    \right)P_{2}(\mu)
    +
    \frac{18\xi^{2}}{35}
    P_{4}(\mu),
    \qquad
    \mu=\cos\theta.
    \label{eq:III_multipoles_exact}
\end{align}
Consequently, to first order in the anisotropy parameter,
\begin{equation}
    \left[
        1+\xi P_{2}(\cos\theta)
    \right]^{2}
    =
    1
    +
    2\xi P_{2}(\cos\theta)
    +
    {\cal O}(\xi^{2}).
    \label{eq:III_multipoles_linear}
\end{equation}
The $\ell=4$ sector is therefore absent at
${\cal O}(\xi)$.  This feature will allow the leading
spin--torsion deformation of the geometry to be organized into a
monopolar sector and an even-parity quadrupolar sector.

From this point onward the paper uses a strict expansion through
$\order(\xi)$.  In particular, the monopole factor
$1+\xi^2/5$ is not retained unless the accompanying $\ell=2$ and
$\ell=4$ terms at the same order are also included.  The results below
therefore carry a remainder $\order(\xi^2)$ and make no claim of an
exact symmetry under $\xi\to-\xi$.

There is one further consistency condition that is essential for the
gravitational perturbation problem.  For an inhomogeneous spin
profile the contact contribution in
Equation~\eqref{eq:III_Uspin} is not separately conserved:
\begin{equation}
    \bar\nabla^{\mu}
    U_{\mu\nu}^{\rm spin}
    =
    \partial_{\nu}{\cal S}
    =
    +\frac{3\kappa}{16}
    \partial_{\nu}J_{5}^{2}.
    \label{eq:III_U_divergence}
\end{equation}
The right-hand side vanishes in the $t$ and $\phi$ directions because
of stationarity and axisymmetry, but it is generically nonzero in the
$r$ and $\theta$ directions.  This does not represent a violation of
energy--momentum conservation.  The decomposition used in
section~\ref{sec:ECDM} separates the explicit four-fermion contact term
from the Dirac kinetic contribution, whereas only their sum is
conserved on the matter equations of motion.

Accordingly, the source entering the linearized gravitational
equations must be a complete conserved effective stress tensor,
\begin{equation}
    \tau_{\mu\nu}^{(T)}
    \equiv
    U_{\mu\nu}^{\rm spin}
    +
    \Theta_{\mu\nu},
    \label{eq:III_tau}
\end{equation}
where $\Theta_{\mu\nu}$ denotes the response of the supporting matter
sector relative to the torsion-free configuration.  The equation
\begin{equation}
    \bar\nabla^{\mu}
    \Theta_{\mu\nu}
    =
    -
    \partial_{\nu}{\cal S},
    \label{eq:III_Theta_conservation}
\end{equation}
so that
\begin{equation}
    {
    \bar\nabla^{\mu}
    \tau_{\mu\nu}^{(T)}
    =
    0
    } .
    \label{eq:III_tau_conservation}
\end{equation}
does not determine $\Theta_{\mu\nu}$ uniquely.  We therefore close the
model rather than treating eq.~\eqref{eq:III_Theta_conservation} as a
definition.

To keep the perturbative normalization explicit, the physical source
in this section and the coefficient source used in the linearized
Einstein equation are related by
\begin{equation}
 \tau_{\mu\nu}^{(T)}=\epsilon_T\widehat\tau_{\mu\nu}^{(T)},
 \qquad \widehat{\cal S}=\epsilon_T^{-1}{\cal S},
 \qquad \bar\nabla^\mu\widehat\tau_{\mu\nu}^{(T)}=0.
 \label{eq:III_normalized_source}
\end{equation}
Thus $\widehat\tau_{\mu\nu}^{(T)}$ is held fixed as
$\epsilon_T\to0$; no power of $\epsilon_T$ is hidden in the
coefficient equations of section~\ref{sec:near_horizon_geometry}.

The closure used in this paper is a phenomenological, compactly
supported anisotropic source.  It is not claimed to follow from a
global solution of the Hehl--Datta equation.  At the order needed for
the leading BZ response, write
\begin{equation}
 {\cal S}(r,\theta)={\cal S}_0(r)
 [1+2\xi P_2(\cos\theta)]+\order(\xi^2),
 \qquad
 {\cal A}_T\equiv2{\cal S}+r\partial_r{\cal S},
 \label{eq:III_completion_SA}
\end{equation}
and define
\begin{equation}
 {\cal D}_T(r,\theta)\equiv-\frac{1}{\sin^2\theta}
 \int_0^\theta\!\dd\vartheta\,\sin^2\vartheta\,
 \partial_\vartheta{\cal A}_T(r,\vartheta).
 \label{eq:III_completion_D}
\end{equation}
In mixed Schwarzschild components the nonrotating completion is
specified completely by
\begin{align}
 \tau^{(T)t}{}_{t}&={\cal S},&
 \tau^{(T)r}{}_{r}&={\cal S},
 \nonumber\\
 \tau^{(T)\theta}{}_{\theta}&=
 \frac12({\cal A}_T+{\cal D}_T),&
 \tau^{(T)\phi}{}_{\phi}&=
 \frac12({\cal A}_T-{\cal D}_T),
 \label{eq:III_completion_components}
\end{align}
with all nonrotating shear components set to zero.  Equivalently,
$\Theta^t{}_t=\Theta^r{}_r=0$ and the two angular components of
$\Theta^\mu{}_\nu$ are obtained by subtracting ${\cal S}$ from
eq.~\eqref{eq:III_completion_components}.  The two nontrivial
conservation equations reduce identically to
\begin{align}
 \partial_r{\cal S}+\frac{2{\cal S}-{
 \tau^{(T)\theta}{}_{\theta}}-
 \tau^{(T)\phi}{}_{\phi}}{r}&=0,
 \nonumber\\
 \partial_\theta\tau^{(T)\theta}{}_{\theta}
 +\cot\theta\left(
 \tau^{(T)\theta}{}_{\theta}-
 \tau^{(T)\phi}{}_{\phi}\right)&=0.
 \label{eq:III_completion_check}
\end{align}
Thus $\bar\nabla_\mu\tau^{(T)\mu}{}_{\nu}=0$ follows by direct
substitution, not by leaving undetermined source functions.  For the
quadrupolar profile one finds explicitly
\begin{align}
 \tau^{(T)\theta}{}_{\theta}
 &=\frac{{\cal A}_0}{2}
 [1+\xi(1+P_2)]+\order(\xi^2),
 \nonumber\\
 \tau^{(T)\phi}{}_{\phi}
 &=\frac{{\cal A}_0}{2}
 [1+\xi(3P_2-1)]+\order(\xi^2),
 \qquad
 {\cal A}_0=2{\cal S}_0+r{\cal S}_0'.
 \label{eq:III_completion_xi}
\end{align}

The rotational part is an additional property of the effective
source and is not fixed by the scalar $J_5^2$.  In dimensionless
coordinates $\hat t=t/M$ and $y=r/M$ we specify its axial component by
\begin{align}
 \kappa\tau^{(T)\hat t}{}_{\phi}
 &=-\frac{2\pi\epsilon_T\chi}{M^2}
 {\mathfrak j}(y)\sin^2\theta
 [1+2\xi P_2(\cos\theta)],
 &
 \kappa\tau^{(T)\phi}{}_{\hat t}
 &=\frac{2\pi\epsilon_T\chi}{M^2}
 \frac{f(y)}{y^2}{\mathfrak j}(y)
 [1+2\xi P_2(\cos\theta)]
 \nonumber\\
 &\quad-\frac{2\chi}{y^3}\kappa
 \left[
 \tau^{(T)\phi}{}_{\phi}
 -\tau^{(T)\hat t}{}_{\hat t}
 \right]_{\chi=0},
 \label{eq:III_completion_rotation}
\end{align}
where $f(y)=1-2/y$ and
\begin{equation}
 {\mathfrak j}(y)=\lambda_J y^{2q}{\cal W}_{\rm NH}^2(y),
 \qquad 2\le y\le2+w,
 \label{eq:III_jprofile}
\end{equation}
and it vanishes outside the source.  The factor
$1+2\xi P_2(\cos\theta)$ is inherited from the contact-stress profile
at the working order; it is essential for closing the
$\order(\epsilon_T\chi\xi)$ axial sector.  In the second line of
eq.~\eqref{eq:III_completion_rotation}, the first term is the
Schwarzschild index-raised axial source, while the final term is the
kinematic correction required when the symmetric covariant tensor is
converted to mixed components with the slow-Kerr metric.  It introduces
no new source function and leaves the independent component
$\tau^{\hat t}{}_{\phi}$ unchanged.  Together the two terms enforce
symmetry of $\tau_{\mu\nu}$ through $\order(\epsilon_T\chi\xi)$; all
$r\phi$, $\theta\phi$, $tr$, and $t\theta$ components vanish at this
order.

The conservation statement must be made on the same background as the
linearized Einstein equation.  Expanding both the connection and the
source about Schwarzschild gives
\begin{equation}
 \left[\bar\nabla_\mu
 \widehat\tau^{\mu}{}_{\nu}\right]_{\chi}
 =\nabla^{(0)}_\mu\widehat\tau^{(1)\mu}{}_{\nu}
 +(\delta\nabla^{(1)}_\mu)
 \widehat\tau^{(0)\mu}{}_{\nu}=0.
 \label{eq:III_slowKerr_conservation}
\end{equation}
For the circular completion adopted here, direct expansion of the
Christoffel symbols shows that the Kerr-connection cross term and the
projected rotating-source divergence vanish separately for every
$\nu=t,r,\theta,\phi$.  The full component derivation and Mathematica
residuals are given in Appendix~\ref{app:slow_kerr_conservation} and
table~\ref{tab:axial_validation}.  The dimensionless parameter $\lambda_J$
is part of the source model.  Its sign distinguishes co-rotating and
counter-rotating supporting matter.  In a local rigidly rotating
spin-fluid realization, let
\begin{equation}
 u^\mu=u^t(\partial_t+\Omega_{\rm m}\partial_\phi)^\mu,
 \qquad S^{\mu\nu}u_\nu=0,
 \label{eq:III_spinfluid_velocity}
\end{equation}
with the Frenkel condition for the Weyssenhoff spin tensor
\cite{Hehl1976,ObukhovKorotky1987}.  The convective angular-momentum
density has the schematic slow-rotation form
\begin{equation}
 -\kappa\tau^{\hat t}{}_{\phi}
 ={\cal L}_{\rm m}(r)(\Omega_{\rm m}-\omega_{\rm K})
 \sin^2\theta+\tau^{\rm spin}_{\hat t\phi},
 \label{eq:III_spinfluid_current}
\end{equation}
where ${\cal L}_{\rm m}>0$ is the relativistic inertia density with
the coordinate normalization included.  Define the dimensionless
coarse-grained current
\begin{equation}
 {\cal J}_{\rm m}(y,\theta)\equiv
 -\frac{\kappa M^2}{\epsilon_T\chi}
 \frac{\tau^{\hat t}{}_{\phi}}{\sin^2\theta}.
 \label{eq:III_lambdaJ_interpretation}
\end{equation}
The closure in eq.~\eqref{eq:III_completion_rotation} is the special
case ${\cal J}_{\rm m}=2\pi\lambda_Jy^{2q}{\cal W}_{\rm NH}^2
[1+2\xi P_2]$.  Equivalently, $\lambda_J$ is the radially constant
$\ell=1$ projection of ${\cal J}_{\rm m}$ after division by the
specified window.  This relation includes both the convective term in
eq.~\eqref{eq:III_spinfluid_current} and the derivative spin stress.
It explains the expected sign and order of magnitude: ordinary
co-rotation gives $\lambda_J>0$, counter-rotation gives
$\lambda_J<0$, and $\lambda_J=\order(1)$ means that the rotational
stress has the natural EC scale $\epsilon_T\chi/(\kappa M^2)$.  It
does not determine a universal number without an equation of state,
spin transport law and matter rotation profile; those are additional
microscopic data of a Weyssenhoff or Dirac-fluid model
\cite{HehlDatta1971,ObukhovKorotky1987,
VignoloDeMariaFabbriCarloni2025,AbboudGavassinoSinghSperanza2025}.

Equations~\eqref{eq:III_completion_components} and
\eqref{eq:III_completion_rotation} are the explicitly specified source
completion used to the perturbative order below.  Other conserved
completions define different effective models and generally give a
different BZ response.  In the minimal theory there is still no
direct torsion term in $F=\dd A$; electromagnetic corrections enter
through the geometry and force-free boundary data.

The construction above therefore provides a controlled hierarchy of
small parameters,
\begin{equation}
    \epsilon_T\ll1,
    \qquad
    |\xi|\ll1,
    \qquad
    \chi^2\ll1,
    \label{eq:III_hierarchy}
\end{equation}
with the slow-rotation expansion introduced only when an analytic
closed form is required.  Through the retained order, the
scalar/even-parity sector contains $\ell=0,2$, whereas the rotational
odd-parity sector contains $\ell=1,3$ at
$\order(\epsilon_T\chi\xi)$.  The source is compactly
supported in the radial interval
$r_{+}\leq r<r_{+}+\Delta r$, and vanishes identically in the exterior
matter vacuum.  These properties make it suitable for the analytic
construction of the stationary axisymmetric metric perturbation in
the next section.

\section{Analytic near-horizon geometry}
\label{sec:near_horizon_geometry}

The perturbation is defined from the outset in a fixed-charge
\emph{comparison ensemble}: after solving the sourced equations we
subtract the homogeneous Kerr mass and angular-momentum variations so
that the perturbed geometry and its reference Kerr solution have the
same ADM charges.  Thus $\delta M_{\rm ADM}=\delta J_{\rm ADM}=0$ are
not merely local boundary conditions; they define the branch being
compared and require a compensating external support or matching
sector.  The explicit subtractions appear below.  This distinction is
the standard separation between sourced perturbations and homogeneous
parameter modes in black-hole perturbation theory
\cite{ReggeWheeler1957,Moncrief1974,MartelPoisson2005,Poisson2004}.

We now determine the stationary and axisymmetric metric response
generated by the spin--torsion source constructed in
section~\ref{sec:spin_source}. The calculation is performed
perturbatively around the Kerr geometry and is organized as a double
expansion in the dimensionless torsion amplitude
$\epsilon_T$ and the dimensionless rotation parameter
\begin{equation}
    \chi
    \equiv
    \frac{a}{M}.
    \label{eq:IV_chi}
\end{equation}
The torsion itself remains algebraically tied to the local axial
current. The physical metric source is the conserved effective stress
defined in eq.~\eqref{eq:III_tau}.  The force-free magnetosphere is
treated in the usual negligible-self-gravity approximation: its
electromagnetic stress transports energy and angular momentum but is
not included as an independent source of the background metric
perturbation considered here.  In the coefficient equation below
we use the normalized tensor of eq.~\eqref{eq:III_normalized_source}:
\begin{equation}
    \widehat\tau_{\mu\nu}^{(T)}
    =\epsilon_T^{-1}(U_{\mu\nu}^{\rm spin}+\Theta_{\mu\nu}),
    \qquad
    \bar\nabla^{\mu}
    \widehat\tau_{\mu\nu}^{(T)}
    =
    0.
    \label{eq:IV_tau_conserved}
\end{equation}
The conservation condition in eq.~\eqref{eq:IV_tau_conserved} is
essential: the explicit contact term
$U_{\mu\nu}^{\rm spin}$ is not separately conserved for an
inhomogeneous axial-current profile and therefore cannot, by itself,
be used consistently as the source of the linearized Einstein
equations.

At the retained slow-rotation order,
$\bar\nabla=\nabla^{(0)}+\chi\delta\nabla^{(1)}+
\order(\chi^2)$ and
$\widehat\tau=\widehat\tau^{(0)}+\chi\widehat\tau^{(1)}+
\order(\chi^2)$.  The complete first-order condition is therefore
eq.~\eqref{eq:III_slowKerr_conservation}, rather than conservation of
the rotating source on Schwarzschild alone.
Appendix~\ref{app:slow_kerr_conservation} evaluates both terms directly from
the slow-Kerr Christoffel symbols and finds four zero residuals for
each term separately through $\order(\xi)$.  This supplies the
order-by-order source identity required by the linearized Bianchi
identity.

We write the full metric as
\begin{equation}
    g_{\mu\nu}
    =
    \bar g_{\mu\nu}
    +
    \epsilon_T h_{\mu\nu}
    +
    {\cal O}(\epsilon_T^2),
    \label{eq:IV_metric_expansion}
\end{equation}
where $\bar g_{\mu\nu}$ is the Kerr metric
eq.~\eqref{eq:III_Kerr}. The first-order field equation is
\begin{equation}
    {
    \delta G_{\mu\nu}[h]
    =
    \kappa\,
    \widehat\tau_{\mu\nu}^{(T)}
    }
    \label{eq:IV_linearized_Einstein}
\end{equation}
with
\begin{align}
    \delta G_{\mu\nu}[h]
    =&\;
    -\frac12
    \bar\nabla^{2}h_{\mu\nu}
    -
    \frac12
    \bar\nabla_{\mu}\bar\nabla_{\nu}h
    +
    \bar\nabla^{\rho}
    \bar\nabla_{(\mu}h_{\nu)\rho}
    \nonumber\\
    &
    -
    \frac12
    \bar g_{\mu\nu}
    \left(
        \bar\nabla_{\rho}\bar\nabla_{\sigma}
        h^{\rho\sigma}
        -
        \bar\nabla^{2}h
    \right)
    -
    \bar R_{\mu\rho\nu\sigma}
    h^{\rho\sigma},
    \label{eq:IV_deltaG}
\end{align}
where
\begin{equation}
    h
    \equiv
    \bar g^{\mu\nu}h_{\mu\nu}.
\end{equation}
Since the Kerr background is Ricci flat,
\begin{equation}
    \bar R_{\mu\nu}=0,
    \qquad
    \bar G_{\mu\nu}=0,
\end{equation}
Equation~\eqref{eq:IV_deltaG} contains no terms proportional to the
background Ricci tensor.  The linearized Bianchi identity,
\begin{equation}
    \bar\nabla^{\mu}
    \delta G_{\mu\nu}[h]
    =
    0,
\end{equation}
is consistent with eq.~\eqref{eq:IV_tau_conserved} by construction.

A completely general stationary and axisymmetric metric perturbation
contains ten independent components,
\begin{equation}
    h_{\mu\nu}
    =
    \begin{pmatrix}
        h_{tt}      & h_{tr}      & h_{t\theta}      & h_{t\phi}
        \\
        h_{tr}      & h_{rr}      & h_{r\theta}      & h_{r\phi}
        \\
        h_{t\theta} & h_{r\theta} & h_{\theta\theta} & h_{\theta\phi}
        \\
        h_{t\phi}   & h_{r\phi}   & h_{\theta\phi}   & h_{\phi\phi}
    \end{pmatrix},
    \label{eq:IV_ten_components}
\end{equation}
with
\begin{equation}
    \partial_t h_{\mu\nu}
    =
    \partial_\phi h_{\mu\nu}
    =
    0.
    \label{eq:IV_stationary_axisymmetric}
\end{equation}
Equatorial reflection symmetry further distinguishes the even and odd
angular sectors. The source constructed in section~\ref{sec:spin_source}
has even parity in the scalar sector, while rotation generates the
axial $t\phi$ response at odd order in $\chi$.

For analytic control we expand
\begin{equation}
    h_{\mu\nu}
    =
    h_{\mu\nu}^{(0)}
    +
    \chi\,h_{\mu\nu}^{(1)}
    +
    \chi^{2}h_{\mu\nu}^{(2)}
    +
    {\cal O}(\chi^{3}),
    \label{eq:IV_slow_rotation_expansion}
\end{equation}
so that the full metric perturbation is
\begin{equation}
    \delta_T g_{\mu\nu}
    =
    \epsilon_T
    \left[
        h_{\mu\nu}^{(0)}
        +
        \chi h_{\mu\nu}^{(1)}
        +
        \chi^2 h_{\mu\nu}^{(2)}
    \right]
    +
    {\cal O}
    \left(
       \epsilon_T\chi^3,
       \epsilon_T^2
    \right).
    \label{eq:IV_double_expansion}
\end{equation}
The normalized coefficient source is expanded in the same way,
\begin{equation}
    \widehat\tau_{\mu\nu}^{(T)}
    =
    \widehat\tau_{\mu\nu}^{(0)}
    +
    \chi\,\widehat\tau_{\mu\nu}^{(1)}
    +
    \chi^{2}\widehat\tau_{\mu\nu}^{(2)}
    +
    {\cal O}(\chi^{3}).
    \label{eq:IV_source_rotation_expansion}
\end{equation}

At linear order in the angular-anisotropy parameter $\xi$, the scalar
spin density derived in section~\ref{sec:spin_source} contains only
$\ell=0$ and $\ell=2$ harmonics,
\begin{equation}
    J_{5}^{2}
    =
    J_{0}^{2}
    {\cal R}(r)
    \left[
        1
        +
        2\xi P_{2}(\cos\theta)
    \right]
    +
    {\cal O}(\xi^{2}),
    \label{eq:IV_J5_multipoles}
\end{equation}
where
\begin{equation}
    {\cal R}(r)
    =
    \left(
        \frac{r}{M}
    \right)^{2q}
    {\cal W}_{\rm NH}^{2}(r).
    \label{eq:IV_radial_profile}
\end{equation}
It is therefore sufficient, at
${\cal O}(\xi)$, to retain the monopole and quadrupole sectors of
the polar perturbation.  We introduce the decomposition
\begin{align}
    h_{tt}
    &=
    H_{tt}^{(0)}(r)
    +
    \xi H_{tt}^{(2)}(r)
    P_{2}(\cos\theta)
    +
    {\cal O}(\xi^{2}),
    \label{eq:IV_htt}
    \\
    h_{rr}
    &=
    H_{rr}^{(0)}(r)
    +
    \xi H_{rr}^{(2)}(r)
    P_{2}(\cos\theta)
    +
    {\cal O}(\xi^{2}),
    \label{eq:IV_hrr}
    \\
    h_{\theta\theta}
    &=
    r^{2}
    \left[
        K^{(0)}(r)
        +
        \xi K^{(2)}(r)
        P_{2}(\cos\theta)
    \right]
    +
    {\cal O}(\xi^{2}),
    \label{eq:IV_hthetatheta}
    \\
    h_{\phi\phi}
    &=
    r^{2}\sin^{2}\theta
    \left[
        K^{(0)}(r)
        +
        \xi K_{\phi}^{(2)}(r)
        P_{2}(\cos\theta)
    \right]
    +
    {\cal O}(\xi^{2}),
    \label{eq:IV_hphiphi}
\end{align}
while the leading rotational sector is written as
\begin{equation}
    h_{t\phi}
    =
    -2\chi M^{2}
    \sin^{2}\theta
    \left[
        \omega_{0}(r)
        +
        \xi\omega_{2}(r)
        P_{2}(\cos\theta)
    \right]
    +
    {\cal O}
    \left(
        \chi^{3},
        \xi^{2}
    \right).
    \label{eq:IV_htphi}
\end{equation}
The remaining components
$h_{tr}$,
$h_{t\theta}$,
$h_{r\theta}$,
$h_{r\phi}$, and
$h_{\theta\phi}$
may be retained in the fully covariant formulation
eq.~\eqref{eq:IV_ten_components}; however, for the stationary,
reflection-symmetric and circular sector considered here they can be
removed by a regular gauge choice together with the circularity
conditions.  We adopt a Regge--Wheeler-type gauge in which
\begin{equation}
    h_{tr}
    =
    h_{t\theta}
    =
    h_{r\theta}
    =
    h_{r\phi}
    =
    h_{\theta\phi}
    =
    0.
    \label{eq:IV_gauge}
\end{equation}
The residual gauge freedom and the invariance of the horizon
observables used below are discussed in
Appendix~\ref{app:metric_perturbation}.

The monopole sector is most conveniently organized in Schwarzschild
coordinates at zeroth order in $\chi$.  Defining
\begin{equation}
    f(r)
    =
    1-\frac{2M}{r},
    \label{eq:IV_f}
\end{equation}
we write
\begin{equation}
    ds_{0}^{2}
    =
    -
    f(r)
    \left[
        1+\epsilon_T H_{0}(r)
    \right]dt^{2}
    +
    \frac{
        1+\epsilon_T H_{2}(r)
    }{
        f(r)
    }dr^{2}
    +
    r^{2}d\Omega^{2},
    \label{eq:IV_monopole_metric}
\end{equation}
where the areal-radius gauge has been used in the monopolar sector.
The conserved effective source is decomposed as
\begin{equation}
    \widehat\tau^{\mu}{}_{\nu}
    =
    {\rm diag}
    \left(
        -\rho_T,
        p_{r,T},
        p_{\perp,T},
        p_{\perp,T}
    \right)
    +
    {\cal O}(\chi,\xi).
    \label{eq:IV_tau_monopole}
\end{equation}
For the completion in eq.~\eqref{eq:III_completion_components}, these
functions are not free:
\begin{equation}
 \rho_T=-\widehat{\cal S}_0,
 \qquad p_{r,T}=\widehat{\cal S}_0,
 \qquad p_{\perp,T}=\widehat{\cal S}_0+\frac{r}{2}\widehat{\cal S}_0'.
 \label{eq:IV_explicit_monopole_source}
\end{equation}
The $tt$ equation can be integrated by introducing an effective mass
function $m_T(r)$,
\begin{equation}
    H_{2}(r)
    =
    \frac{
        2m_T(r)
    }{
        r-2M
    },
    \label{eq:IV_H2_mass}
\end{equation}
with
\begin{equation}
    {
    m_T'(r)
    =
    4\pi r^{2}\rho_T(r)
    }.
    \label{eq:IV_mass_equation}
\end{equation}
Thus
\begin{equation}
    m_T(r)
    =
    4\pi
    \int_{r_{+}}^{r}
    dr'\,
    r'^{\,2}
    \rho_T(r')
    +
    C_M.
    \label{eq:IV_mass_solution}
\end{equation}
The integration constant $C_M$ represents a homogeneous mass
perturbation.  Since $M$ is defined as the ADM mass of the background
black hole, we impose
\begin{equation}
    \delta M_{\rm ADM}=0,
    \label{eq:IV_ADMmassfix}
\end{equation}
which removes the corresponding asymptotic homogeneous mode.
Explicitly, the fixed-ADM prescription is
\begin{equation}
 C_M=-4\pi\int_{r_+}^{\infty}\!\dd r\,r^2\rho_T(r),
 \qquad
 m_T(r)=-4\pi\int_r^{\infty}\!\dd r'\,r'^2\rho_T(r').
 \label{eq:IV_fixedADM_mass_explicit}
\end{equation}
This is an ensemble definition: the homogeneous variation
$-\delta M_{\rm src}\,\partial_M g^{\rm K}_{\mu\nu}$ is included so
that the perturbed solution and the reference Kerr black hole have
the same measured ADM mass. It is not a claim that a localized
positive-energy source can be added while the bare Kerr parameters
are held fixed without compensation.

The difference between the $rr$ and $tt$ equations gives the
redshift potential,
\begin{equation}
    H_{0}'(r)
    =
    \frac{
        2m_T(r)
    }{
        r(r-2M)
    }
    +
    \frac{
        8\pi r^{2}p_{r,T}(r)
    }{
        r-2M
    },
    \label{eq:IV_H0_equation}
\end{equation}
and hence
\begin{equation}
    H_{0}(r)
    =
    -
    \int_{r}^{\infty}
    dr'\,
    \left[
        \frac{
            2m_T(r')
        }{
            r'(r'-2M)
        }
        +
        \frac{
            8\pi r'^{\,2}p_{r,T}(r')
        }{
            r'-2M
        }
    \right],
    \label{eq:IV_H0_solution}
\end{equation}
where asymptotic flatness has been used to fix the additive constant.
Equations~\eqref{eq:IV_mass_solution} and
\eqref{eq:IV_H0_solution} provide a closed integral representation of
the full monopolar geometry for an arbitrary conserved
spin--torsion source.

The polar quadrupole sector can be reduced to a single second-order
master equation.  Introducing a gauge-invariant radial master
variable $Z_{2}(r)$, defined explicitly in
Appendix~\ref{app:metric_perturbation}, the $\ell=2$ Einstein system
takes the form
\begin{equation}
    {
    \frac{d^{2}Z_{2}}{dr_{*}^{2}}
    -
    V_{2}^{\rm stat}(r)
    Z_{2}
    =
    {\cal S}_{2}(r)
    }
    \label{eq:IV_Zerilli_static}
\end{equation}
with tortoise coordinate
\begin{equation}
    \frac{dr_{*}}{dr}
    =
    \frac{1}{f(r)}.
    \label{eq:IV_tortoise}
\end{equation}
For $\ell=2$ the static Zerilli potential is
\begin{equation}
    V_{2}^{\rm stat}(r)
    =
    \frac{
        2f(r)
    }{
        r^{3}(2r+3M)^{2}
    }
    \left[
        12r^{3}
        +
        12Mr^{2}
        +
        18M^{2}r
        +
        9M^{3}
    \right].
    \label{eq:IV_Zerilli_potential}
\end{equation}
The source ${\cal S}_{2}(r)$ is the fixed linear combination of the
$\ell=2$ projections of the normalized stress tensor.
Substitution of eq.~\eqref{eq:III_completion_xi} determines every
projection; no independent radial source function remains.  For
reproducibility, Appendix~\ref{app:metric_perturbation} also gives an
equivalent direct Regge--Wheeler-gauge reduction to one explicit
second-order equation for $K^{(2)}$, its Green kernel, and algebraic
reconstruction formulae for $H_0^{(2)}$ and $H_2^{(2)}$.

Let $Z_{2}^{H}(r)$ denote the homogeneous solution regular at the
future horizon and $Z_{2}^{\infty}(r)$ the homogeneous solution
decaying at spatial infinity.  The unique solution satisfying both
physical boundary conditions is
\begin{align}
    Z_{2}(r)
    =&\;
    Z_{2}^{\infty}(r)
    \int_{r_{+}}^{r}
    dr'\,
    \frac{
        Z_{2}^{H}(r')
        {\cal S}_{2}(r')
    }{
        f(r'){\cal W}_{2}
    }
    \nonumber\\
    &
    +
    Z_{2}^{H}(r)
    \int_{r}^{\infty}
    dr'\,
    \frac{
        Z_{2}^{\infty}(r')
        {\cal S}_{2}(r')
    }{
        f(r'){\cal W}_{2}
    },
    \label{eq:IV_Zerilli_green}
\end{align}
where
\begin{equation}
    {\cal W}_{2}
    =
    Z_{2}^{H}
    \frac{dZ_{2}^{\infty}}{dr_{*}}
    -
    Z_{2}^{\infty}
    \frac{dZ_{2}^{H}}{dr_{*}}
    \label{eq:IV_Wronskian}
\end{equation}
is the constant Wronskian.  Equation~\eqref{eq:IV_Zerilli_green} is
an analytic Green-function representation and remains valid for an
arbitrary smooth compactly supported near-horizon source.

The direct reconstruction in
eqs.~\eqref{eq:appIV_K_equation}--\eqref{eq:appIV_H_reconstruction}
is the form used below.  It is algebraically equivalent to the master
description but avoids leaving the source reconstruction hidden in
symbolic differential operators.  In particular, it determines the
Maxwell-driving combination $\beta_2$ in
eq.~\eqref{eq:appIV_beta_explicit} as an explicit functional of the
single prescribed radial profile.

Rotation first generates an axial dipolar frame-dragging response,
while anisotropy also generates an axial octupole.  Through
${\cal O}(\epsilon_T\chi\xi)$ we parameterize
\begin{equation}
    \delta_T g_{t\phi}
    =
    -2\epsilon_T\chi r^{2}
    \left[
      \varpi_1(y)X_\phi^{10}(\theta)
      +\varpi_3(y)X_\phi^{30}(\theta)
    \right]/M
    +
    {\cal O}
    \left(
        \epsilon_T\chi\xi^2,
        \epsilon_T\chi^{3}
    \right).
    \label{eq:IV_frame_dragging_metric}
\end{equation}
For the $\ell=1$ sector, writing $\varpi_1=M\omega_1$, direct symbolic
evaluation of the mixed Einstein component gives
$\delta G^{t}{}_{\phi}
=-\epsilon_T\chi\sin^2\theta\,(r^4\omega_T')'/r^2$.
With the projected source normalized as in
eq.~\eqref{eq:III_completion_rotation}, the field equation is
\begin{equation}
    \frac{d}{dy}
    \left[
        y^{4}
        \frac{d\varpi_1}{dy}
    \right]
    =2\pi\left(1-\frac{2\xi}{5}\right)y^{2}{\mathfrak j}(y),
    \label{eq:IV_omega_equation}
\end{equation}
where ${\mathfrak j}$ is the explicitly prescribed function in
eq.~\eqref{eq:III_jprofile}.  The boundary conditions
\begin{equation}
 \varpi_1(\infty)=0,
 \qquad
 \lim_{y\to\infty}y^4\varpi_1'(y)=0
 \label{eq:IV_omega_boundary}
\end{equation}
The first condition chooses a nonrotating frame at infinity; the
second is the fixed-ADM subtraction of the source-generated exterior
$y^{-3}$ angular-momentum tail. Thus
\begin{equation}
    \delta J_{\rm ADM}=0.
    \label{eq:IV_ADMJfix}
\end{equation}
As in the mass sector, this means that the homogeneous variation
$-\delta J_{\rm src}\,\partial_Jg^{\rm K}_{\mu\nu}$ is part of the
fixed-ADM comparison.  The unique solution gives the horizon value
\begin{equation}
 \varpi_{1H}=\frac{2\pi}{3}
 \left(1-\frac{2\xi}{5}\right)
 \int_2^{2+w}\!\dd y\,{\mathfrak j}(y)
 \left(\frac{y^2}{8}-\frac1y\right).
 \label{eq:IV_varpiH}
\end{equation}
For reference, the fixed-ADM boundary-value problem can be solved
without an indefinite integral.  If
 $F_1(y)=2\pi(1-2\xi/5)\int_2^y\dd z\,z^2{\mathfrak j}(z)$,
 then the coefficient
of the exterior $y^{-3}$ tail is
 $F_1(2+w)+C_1$.  The condition $\delta J_{\rm ADM}=0$ sets
 $C_1=-F_1(2+w)$, and direct integration gives
\begin{equation}
 \varpi_1(y)=\frac{2\pi}{3}\left(1-\frac{2\xi}{5}\right)
 \int_{\max(y,2)}^{2+w}\!\dd z\,{\mathfrak j}(z)
 \left(\frac{z^2}{y^3}-\frac1z\right)
 \label{eq:IV_varpi_solution}
\end{equation}
 inside the source, with $\varpi_1=0$ for $y\ge2+w$; its horizon limit
is eq.~\eqref{eq:IV_varpiH}.  This explicitly exhibits the
compensating fixed-ADM subtraction.
Directly expanding $\Omega_H=-g_{t\phi}/g_{\phi\phi}$ at $r=2M$
gives $\delta\Omega_H=2\epsilon_T\chi\varpi_{1H}/M$, while
$\Omega_H^{\rm K}=\chi/(4M)+\order(\chi^3)$.  The controlled
coefficient is therefore
\begin{align}
 C_\Omega^{(0)}+\xi C_\Omega^{(1)}
 &=8\varpi_{1H}
 =\frac{16\pi\lambda_J}{3}
 \left(1-\frac{2\xi}{5}\right){\cal I}_J(q,w),
 \nonumber\\
 {\cal I}_J(q,w)&=\int_2^{2+w}\!\dd y\,
 y^{2q}{\cal W}_{\rm NH}^2(y)
 \left(\frac{y^2}{8}-\frac1y\right).
 \label{eq:IV_COmega_explicit_completion}
\end{align}
Thus
\begin{equation}
 C_\Omega^{(0)}=\frac{16\pi\lambda_J}{3}{\cal I}_J,
 \qquad C_\Omega^{(1)}=-\frac25C_\Omega^{(0)}.
 \label{eq:IV_COmega_linearxi}
\end{equation}
This result depends explicitly on the rotational completion through
$\lambda_J$; it is not determined by $J_5^2$ alone.

The angular statement can now be made precisely.  With the
axisymmetric odd-parity vector harmonics
$X_\phi^{\ell0}=-\sin\theta\,\partial_\theta P_\ell(\cos\theta)$,
direct projection gives
\begin{equation}
 \sin^2\theta P_2(\cos\theta)
 =-\frac15X_\phi^{10}+\frac15X_\phi^{30},
 \qquad
 \sin^2\theta[1+2\xi P_2]
 =\left(1-\frac{2\xi}{5}\right)X_\phi^{10}
 +\frac{2\xi}{5}X_\phi^{30}.
 \label{eq:IV_axial_decomposition}
\end{equation}
There is therefore an $\ell=1$ projection at
$\order(\epsilon_T\chi\xi)$, and it is exactly the origin of
$C_\Omega^{(1)}$ above.  The remaining $\ell=3$ radial amplitude
$\varpi_3(y)$ obeys
\begin{equation}
 {\cal L}_3[\varpi_3]
 \equiv\frac{d}{dy}\left(y^4\frac{d\varpi_3}{dy}\right)
 -\frac{10y^3}{y-2}\varpi_3
 =\frac{4\pi\xi}{5}y^2{\mathfrak j}(y),
 \label{eq:IV_omega3_equation}
\end{equation}
so the formerly unspecified eigenvalue is
$\lambda_3=(\ell-1)(\ell+2)=10$.  Its physical solution is
\begin{equation}
 \varpi_3(y)=\int_2^\infty\!dz\,G_3(y,z)
 \frac{4\pi\xi}{5}z^2{\mathfrak j}(z),
 \label{eq:IV_omega3_green}
\end{equation}
with the explicit Green kernel in
Appendix~\ref{app:metric_perturbation}.  The horizon-regular
homogeneous solution is proportional to $(y-2)(3y-4)$, hence
\begin{equation}
 \varpi_3(2)=0.
 \label{eq:IV_omega3_horizon}
\end{equation}
The $\ell=3$ sector is therefore nonzero away from the horizon but
does not make the horizon angular velocity latitude dependent or add
another term to $C_\Omega^{(1)}$.

The boundary conditions imposed on the perturbation are therefore
\begin{align}
    &h_{\mu\nu}
    \;\text{regular on the future event horizon},
    \label{eq:IV_BC_horizon}
    \\
    &h_{\mu\nu}
    \rightarrow0
    \qquad
    (r\rightarrow\infty),
    \label{eq:IV_BC_infinity}
    \\
    &\delta M_{\rm ADM}=0,
    \qquad
    \delta J_{\rm ADM}=0.
    \label{eq:IV_BC_ADM}
\end{align}
Regularity at the horizon is imposed in ingoing Kerr coordinates
rather than in Boyer--Lindquist coordinates.  This distinction is
necessary because individual Boyer--Lindquist components may diverge
at $\Delta=0$ even when the corresponding tensor is regular.

The perturbed event horizon is defined by the null surface
\begin{equation}
    {\cal H}:
    \qquad
    r
    =
    r_{+}
    +
    \epsilon_T\,
    \delta r_{H}(\theta).
    \label{eq:IV_horizon_surface}
\end{equation}
Let
\begin{equation}
    \chi^{\mu}_{H}
    =
    t^{\mu}
    +
    \Omega_{H}\phi^{\mu}
    \label{eq:IV_horizon_generator}
\end{equation}
be the horizon-generating Killing field.  The null condition
\begin{equation}
    g_{\mu\nu}
    \chi_{H}^{\mu}
    \chi_{H}^{\nu}
    \big|_{\cal H}
    =
    0
    \label{eq:IV_horizon_null}
\end{equation}
determines the first-order horizon shift.  Writing
\begin{equation}
    \Omega_{H}
    =
    \bar\Omega_{H}
    +
    \epsilon_T\,
    \delta\Omega_{H},
    \label{eq:IV_Omega_expand}
\end{equation}
the perturbation of the horizon angular velocity can be expressed in
the gauge-invariant form
\begin{align}
    \delta\Omega_{H}
    =
    -
    \left.
    \frac{
        h_{t\phi}
        +
        \bar\Omega_{H}h_{\phi\phi}
        +
        \delta r_{H}
        \partial_{r}
        \left(
            \bar g_{t\phi}
            +
            \bar\Omega_{H}
            \bar g_{\phi\phi}
        \right)
    }{
        \bar g_{\phi\phi}
    }
    \right|_{r=r_{+}} .
    \label{eq:IV_deltaOmega}
\end{align}
Equivalently, one may use
\begin{equation}
    \Omega_H
    =
    -
    \left.
    \frac{
        g_{t\phi}
    }{
        g_{\phi\phi}
    }
    \right|_{\cal H},
    \label{eq:IV_Omega_ratio}
\end{equation}
provided the horizon location is perturbed consistently.

It is useful to define the dimensionless horizon-response
coefficients
\begin{equation}
    C_{r}
    \equiv
    \frac{
        \delta r_{H}
    }{
        M
    },
    \qquad
    C_{\Omega}
    \equiv
    \frac{
        \delta\Omega_{H}
    }{
        \bar\Omega_{H}
    },
    \label{eq:IV_CrCOmega}
\end{equation}
so that
\begin{equation}
    r_{H}
    =
    r_{+}
    +
    \epsilon_T M C_{r},
    \qquad
    \Omega_{H}
    =
    \bar\Omega_{H}
    \left(
       1+\epsilon_T C_{\Omega}
    \right).
    \label{eq:IV_horizon_responses}
\end{equation}
Both coefficients are linear functionals of the conserved
spin--torsion source.

The horizon area is
\begin{equation}
    A_{H}
    =
    \int_{\cal H}
    d\theta\,d\phi\,
    \sqrt{
        g_{\theta\theta}
        g_{\phi\phi}
        -
        g_{\theta\phi}^{2}
    },
    \label{eq:IV_area_definition}
\end{equation}
which gives
\begin{equation}
    A_{H}
    =
    \bar A_H
    \left[
        1
        +
        \epsilon_T C_A
    \right],
    \qquad
    \bar A_H
    =
    4\pi
    \left(
        r_{+}^{2}+a^{2}
    \right),
    \label{eq:IV_area_response}
\end{equation}
with
\begin{align}
    C_A
    =
    \frac{1}{2\bar A_H}
    \int_{0}^{2\pi}d\phi
    \int_{0}^{\pi}d\theta\,
    \sqrt{
        \bar g_{\theta\theta}
        \bar g_{\phi\phi}
    }
    \left[
        \frac{
            h_{\theta\theta}
        }{
            \bar g_{\theta\theta}
        }
        +
        \frac{
            h_{\phi\phi}
        }{
            \bar g_{\phi\phi}
        }
    \right]_{r=r_+}
    +
    C_A^{(r)},
    \label{eq:IV_CA}
\end{align}
where $C_A^{(r)}$ denotes the contribution from the perturbation of
the horizon location.

The surface gravity may likewise be written as
\begin{equation}
    \kappa_H
    =
    \bar\kappa_H
    \left(
        1+\epsilon_T C_{\kappa}
    \right),
    \label{eq:IV_surface_gravity_response}
\end{equation}
where
\begin{equation}
    \bar\kappa_H
    =
    \frac{
        r_{+}-r_{-}
    }{
        2(r_{+}^{2}+a^{2})
    }.
    \label{eq:IV_kappa_Kerr}
\end{equation}
The coefficient $C_\kappa$ is defined invariantly by
\begin{equation}
    \kappa_H^{2}
    =
    -
    \frac12
    \left(
        \nabla_{\mu}\chi_{H,\nu}
    \right)
    \left(
        \nabla^{\mu}\chi_H^{\nu}
    \right)
    \bigg|_{\cal H}
    \label{eq:IV_surface_gravity_definition}
\end{equation}
and is a linear functional of the even-parity metric response.  Since
it does not enter the controlled leading BZ coefficient, we do not
quote a partially resummed closed form for it.

The construction above yields the near-horizon geometry in a form
that is completely determined by the conserved spin--torsion source.
Schematically,
\begin{equation}
    {
    h_{\mu\nu}(x)
    =
    \kappa
    \int d^{4}x'\,
    G_{\mu\nu}{}^{\rho\sigma}(x,x')
    \tau_{\rho\sigma}^{(T)}(x')
    }
    \label{eq:IV_master_green}
\end{equation}
where the Green tensor is understood with future-horizon regularity,
asymptotic flatness, and fixed ADM mass and angular momentum.
Because the source has compact radial support, the exterior
$r>r_{+}+\Delta r$ is torsion-free but is not, in general,
identical to the unperturbed Kerr spacetime: it contains the
gravitational multipoles induced by the interior source.

The perturbative hierarchy requires
\begin{equation}
    \max_{\mathcal D_{\rm NH}}
    \left|
       \epsilon_T
       \frac{
           h_{\mu\nu}
       }{
           \bar g_{\mu\nu}
       }
    \right|
    \ll1
    \label{eq:IV_metric_validity}
\end{equation}
for every nonvanishing background component, together with
\begin{equation}
    |\epsilon_T C_r|
    \ll1,
    \qquad
    |\epsilon_T C_\Omega|
    \ll1,
    \qquad
    |\epsilon_T C_A|
    \ll1.
    \label{eq:IV_horizon_validity}
\end{equation}
Under these conditions the spin--torsion geometry constitutes a
controlled deformation of Kerr.  The quantities
$\delta\Omega_H$,
$\delta A_H$, and the horizon values of the reconstructed metric
functions provide precisely the geometric input required for the
force-free magnetosphere and Blandford--Znajek calculation in the
next section.

\begin{figure}[t]
  \centering
  \includegraphics[width=.82\textwidth]{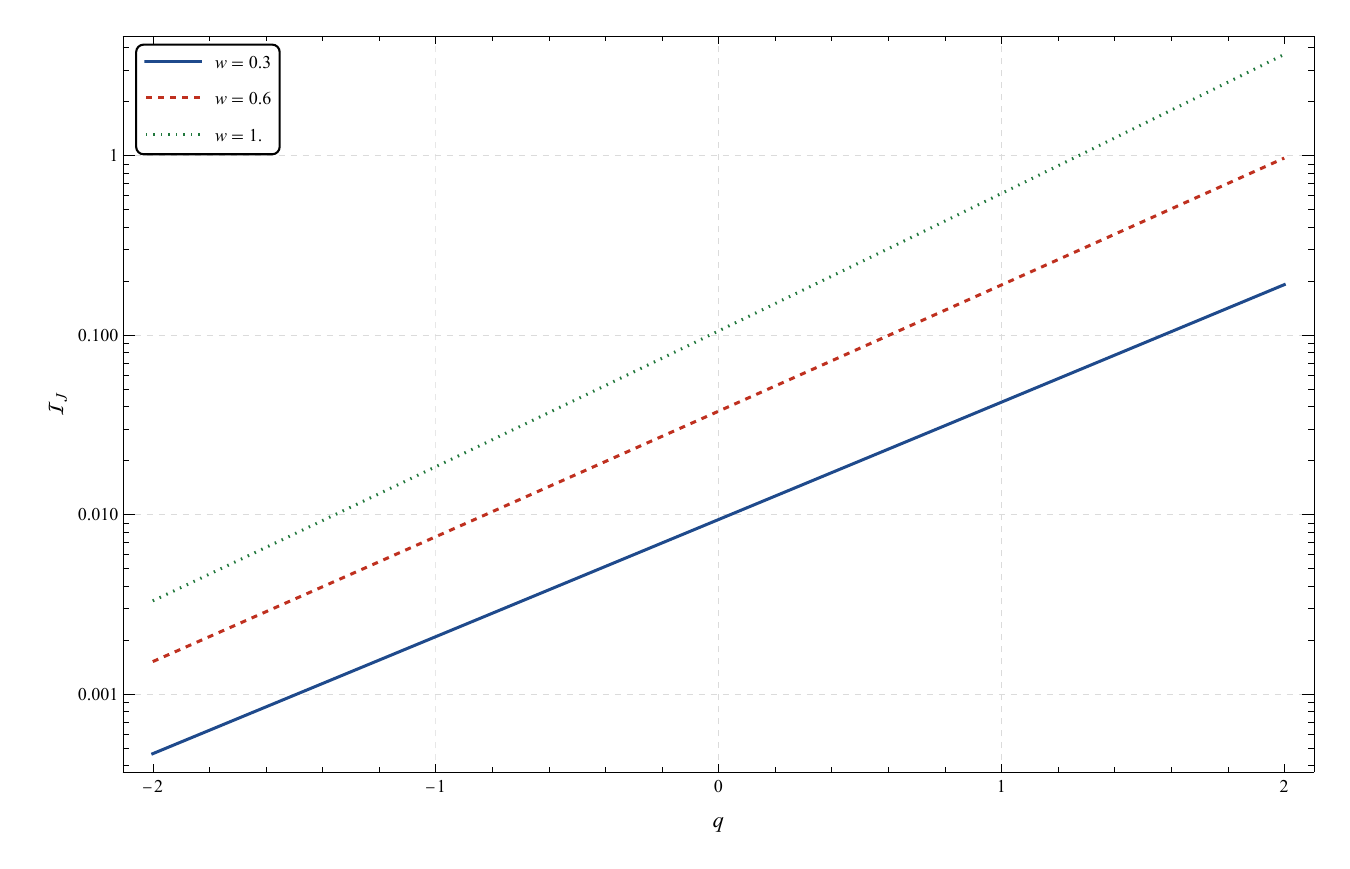}
  \caption{Positive rotational source moment ${\cal I}_J(q,w)$ for
  three compact-support widths.  These curves are direct quadratures
  of eq.~\eqref{eq:IV_IJ}; multiplying by
  $(16\pi\lambda_J/3)(1-2\xi/5)$ gives the leading relative
  horizon-angular-velocity response. The ordinate is logarithmic so that all three
  widths remain visible over the displayed range of $q$.}
  \label{fig:IV_source_moment}
\end{figure}

The result needed below is the explicitly sourced leading
frame-dragging response in eq.~\eqref{eq:IV_COmega_explicit_completion}.
It is useful to isolate its positive radial moment,
\begin{equation}
 {\cal I}_J(q,w)=\int_2^{2+w}\!\dd y\,
 y^{2q}{\cal W}_{\rm NH}^2(y)
 \left(\frac{y^2}{8}-\frac1y\right).
 \label{eq:IV_IJ}
\end{equation}
For $w>0$ the integrand is nonnegative and is strictly positive away
from the horizon endpoint.  Consequently
\begin{equation}
 \operatorname{sgn}[C_\Omega^{(0)}+\xi C_\Omega^{(1)}]
 =\operatorname{sgn}\lambda_J,
 \qquad
 C_\Omega^{(0)}+\xi C_\Omega^{(1)}
 =\frac{16\pi\lambda_J}{3}\left(1-\frac{2\xi}{5}\right)
 {\cal I}_J(q,w).
 \label{eq:IV_COmega_moment}
\end{equation}
The sign equality assumes the physical range $|\xi|\ll1$, for which
$1-2\xi/5>0$. No angular factor of order $\xi^2$ is retained.
Figure~\ref{fig:IV_source_moment} shows the radial moment itself and
therefore separates the calculable radial dependence from the model
parameter $\lambda_J$.

\section{Force-free magnetosphere and the generalized Znajek condition}
\label{sec:force_free}

We use the standard stationary, axisymmetric force-free formalism and
horizon regularity condition
\cite{GoldreichJulian1969,Michel1973,BlandfordZnajek1977,Znajek1977,
MacdonaldThorne1982,GrallaJacobson2014,KoideNodaTakahashi2025,
MeringoloCamilloniRezzolla2025,FengCaiYangZhang2026}.

The exterior plasma is described by stationary, axisymmetric
force-free electrodynamics,
\begin{equation}
 \nabla_\mu F^{\mu\nu}=4\pi J^\nu,
 \qquad F_{\mu\nu}J^\nu=0,
 \qquad \dd F=0.
 \label{eq:V_FFE}
\end{equation}
Here $J^\mu\equiv J_{\rm FF}^\mu$ is the current of the exterior
force-free plasma.  It is physically and mathematically distinct from
the Dirac vector current $J_D^\mu$, which vanishes for the neutral
torsion-producing source by eq.~\eqref{eq:neutral_spin_source}.
In addition to eq.~\eqref{eq:V_FFE}, magnetic domination and
degeneracy require
\begin{equation}
 F_{\mu\nu}{}^\star F^{\mu\nu}=0,
 \qquad
 F_{\mu\nu}F^{\mu\nu}>0,
 \label{eq:V_degeneracy}
\end{equation}
where the second inequality is imposed in the force-free domain away
from current sheets. Stationarity and axisymmetry mean
$\mathcal L_tF=\mathcal L_\phi F=0$. Together with $\dd F=0$, these
conditions introduce the magnetic-flux function $\Psi=A_\phi$ and
imply Ferraro isorotation,
\begin{equation}
 \Omega_F=\Omega_F(\Psi).
 \label{eq:V_Ferraro}
\end{equation}
We define the poloidal current by the sign convention
\begin{equation}
 I(\Psi)=2\pi\sqrt{-g}\,F^{\theta r}.
 \label{eq:V_current_definition}
\end{equation}
The force-free equation then gives $I=I(\Psi)$. With these
conventions the field strength can be written as
\begin{equation}
 F=\dd\Psi\wedge(\dd\phi-\Omega_F\dd t)
   +F_{r\theta}\,\dd r\wedge\dd\theta.
 \label{eq:V_F_decomposition}
\end{equation}
The background is the split monopole
\begin{equation}
 \Psi_0=\Psi_\star(1-\cos\theta).
 \label{eq:V_split_monopole}
\end{equation}
Here $2\pi\Psi_\star$ is the magnetic flux through one hemisphere in
our normalization. The current sheet needed to reverse the monopole
across the equatorial plane is not resolved by the near-horizon
expansion. All equations below are therefore understood separately
in the two smooth hemispheres and matched across that sheet.

Minimal coupling is important here: torsion has already been removed
algebraically, and there is no additional term in $F=\dd A$. The
electromagnetic response is induced by the perturbed metric and its
regularity surfaces.

The static anisotropic response can be closed before addressing the
rotating two-light-surface problem.  In the Regge--Wheeler gauge of
Appendix~\ref{app:metric_perturbation}, write the polar quadrupole as
\begin{align}
 h_{tt}^{(2)}&=fH_0^{(2)}P_2,&
 h_{rr}^{(2)}&=f^{-1}H_2^{(2)}P_2,
 \nonumber\\
 h_{\theta\theta}^{(2)}&=r^2K^{(2)}P_2,&
 h_{\phi\phi}^{(2)}&=r^2\sin^2\theta K^{(2)}P_2,
 \label{eq:V_static_polar_metric}
\end{align}
and define the metric combination
\begin{equation}
 \beta_2(y)\equiv
 \frac12\left[-H_0^{(2)}(y)+H_2^{(2)}(y)\right]-K^{(2)}(y).
 \label{eq:V_beta2}
\end{equation}
Here the relative perturbations of the radial and angular Maxwell
coefficients are, respectively,
\begin{equation}
 \alpha_2=-\frac12(H_0^{(2)}+H_2^{(2)}),
 \qquad
 \beta_2=\frac12(-H_0^{(2)}+H_2^{(2)})-K^{(2)}.
 \label{eq:V_Maxwell_metric_coefficients}
\end{equation}
The first combination multiplies $\partial_r\Psi_0=0$ and hence does
not source the static response, whereas the second multiplies the
nonzero angular derivative of the monopolar flux.  The direct Einstein
reconstruction gives
\begin{equation}
 \beta_2(y)=-K^{(2)}(y)-\frac{y^2}{2}a(y),
 \qquad
 K^{(2)}(y)=\int_2^\infty\!dz\,G_K(y,z){\cal S}_K(z),
 \label{eq:V_beta_explicit_source}
\end{equation}
with every quantity on the right-hand side printed explicitly in
eqs.~\eqref{eq:appIV_sa}--\eqref{eq:appIV_K_Green}.  Thus the
magnetostatic source contains no unassigned polar metric function.
For $\Omega_F=I=0$, and separately in either smooth hemisphere away
from the split-monopole current sheet, the $\phi$ component of
Maxwell's equation is
\begin{equation}
 \partial_r\!\left(\sqrt{-g}\,g^{rr}g^{\phi\phi}
 \partial_r\Psi\right)
 +\partial_\theta\!\left(\sqrt{-g}\,g^{\theta\theta}g^{\phi\phi}
 \partial_\theta\Psi\right)=0.
 \label{eq:V_magnetostatic_Maxwell}
\end{equation}
There is no $4\pi J_D^\phi$ term on the right-hand side because
$q_{\rm spin}=0$.  The equation therefore isolates the deformation of
the poloidal magnetic flux produced by the EC-induced metric response;
the independent force-free plasma current re-enters the rotating
Grad--Shafranov problem below.
Expanding eq.~\eqref{eq:V_magnetostatic_Maxwell} around Schwarzschild
shows explicitly that the quadrupolar metric changes the operator at
$\order(\epsilon_T\xi)$.  With
\begin{equation}
 X_\phi^{20}=-\sin\theta\,\partial_\theta P_2(\cos\theta),
 \qquad
 \Psi_1^{(\xi)}=\xi\Psi_\star R_2(y)X_\phi^{20},
 \label{eq:V_static_flux_ansatz}
\end{equation}
the complete static equation separates as
\begin{equation}
 \frac{d}{dy}\left[f(y)\frac{dR_2}{dy}\right]
 -\frac{6}{y^2}R_2=\frac{\beta_2(y)}{y^2},
 \qquad f(y)=1-\frac2y.
 \label{eq:V_static_flux_radial}
\end{equation}
The mode $X_\phi^{20}$ vanishes both on the axis and at the equator.
It therefore preserves the fixed hemispheric flux while allowing a
nonzero angular redistribution.

Two homogeneous solutions adapted to the physical boundaries are
\begin{align}
 u_H(y)&=y^2(2y-3),
 \nonumber\\
 u_\infty(y)&=1+3y-6y^2
 +\frac32(3-2y)y^2\ln\!\left(\frac{y-2}{y}\right),
 \label{eq:V_static_flux_homogeneous}
\end{align}
where $u_H$ is regular at $y=2$ and $u_\infty$ decays at infinity.
Their weighted Wronskian is
\begin{equation}
 f(y)[u_Hu_\infty'-u_\infty u_H']=-12.
 \label{eq:V_static_flux_Wronskian}
\end{equation}
Consequently the required Green kernel and the unique magnetostatic
response are
\begin{align}
 G_M(y,z)&=-\frac1{12}u_H(y_<)u_\infty(y_>),
 \nonumber\\
 R_2(y)&=\int_2^\infty\!dz\,G_M(y,z)\frac{\beta_2(z)}{z^2},
 \label{eq:V_static_flux_Green}
\end{align}
with $y_<=\min(y,z)$ and $y_>=\max(y,z)$.  In particular,
\begin{equation}
 R_{2H}\equiv R_2(2)
 =-\frac13\int_2^\infty\!dz\,
 \frac{u_\infty(z)\beta_2(z)}{z^2}.
 \label{eq:V_R2H}
\end{equation}
Because $\beta_2$ is reconstructed explicitly from
eqs.~\eqref{eq:appIV_sa}--\eqref{eq:appIV_beta_explicit},
eqs.~\eqref{eq:V_static_flux_Green}--\eqref{eq:V_R2H} are a nested
analytic Green-functional of the stated source; no load or
light-surface datum enters this static sector.

\begin{figure}[t]
 \centering
 \includegraphics[width=\textwidth]{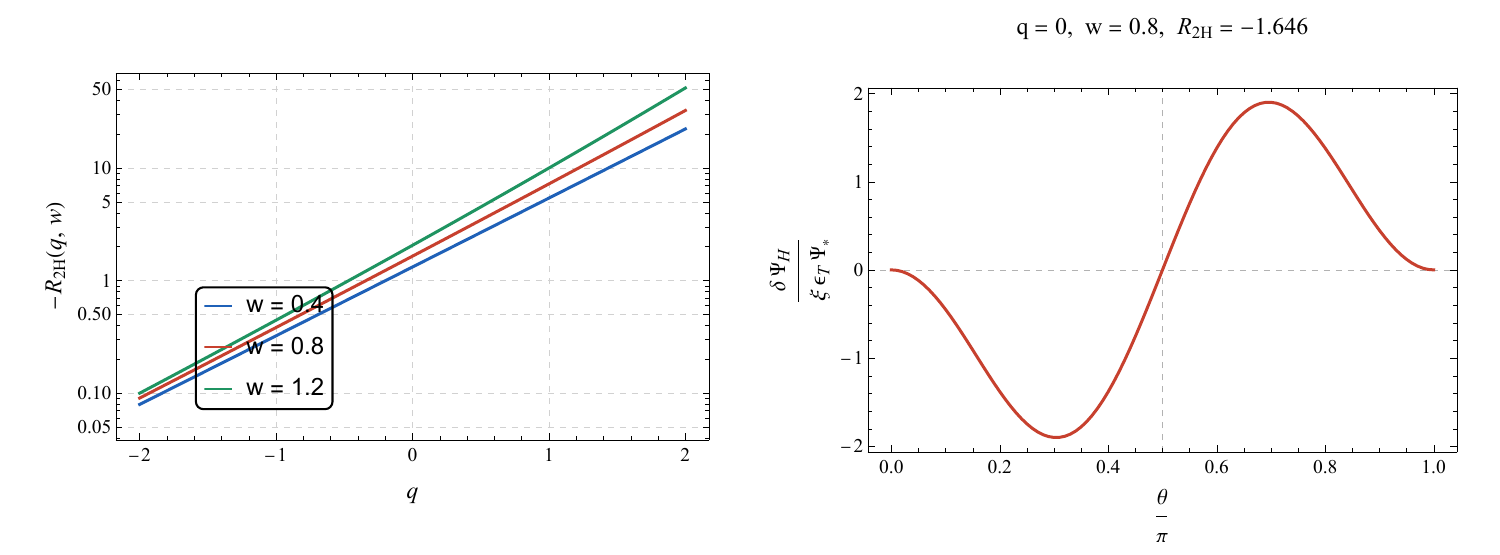}
 \caption{Static magnetostatic response obtained from the nested
 Green quadratures. Left: $-R_{2H}(q,w)$ for three support widths; the
 logarithmic ordinate makes the full displayed range visible.  The
 sampled values have $R_{2H}<0$, but this plot is not used as a general
 sign proof. Right: the normalized horizon redistribution
 $\delta\Psi_H/(\epsilon_T\xi\Psi_\star)=R_{2H}X_\phi^{20}$ for
 $q=0$, $w=0.8$.  Its zeros at the pole and equator preserve the total
 hemispheric flux while its derivative contributes to the BZ power.}
 \label{fig:V_static_flux_response}
\end{figure}

Figure~\ref{fig:V_static_flux_response} displays direct Mathematica
quadratures of this analytic functional.  For the representative
choice $q=0$, $w=0.8$,
\begin{equation}
 R_{2H}=-1.645765886,
 \qquad C_\Phi^{(1)}=1.974919063.
 \label{eq:V_representative_R2H}
\end{equation}
These numbers illustrate the coefficient and are not an independent
simulation or fit.

We now linearize the stream equation and state the boundary data that
make the response problem well posed.

We expand
\begin{equation}
 \Psi=\Psi_0+\epsilon_T\Psi_1,
 \quad I=I_0+\epsilon_T I_1,
 \quad \Omega_F=\Omega_F^{(0)}+\epsilon_T\Omega_F^{(1)}.
 \label{eq:V_FFE_expansion}
\end{equation}
The three first-order functions are not independent. The toroidal
Maxwell equation supplies the stream equation, whereas regularity at
the inner and outer light surfaces constrains $I_1(\Psi_0)$ and
$\Omega_F^{(1)}(\Psi_0)$. Before imposing those constraints, the
linearized Grad--Shafranov equation has the operator form
\begin{equation}
 {\cal L}_{\rm K}[\Psi_1]
 =S_T[h_{\mu\nu}^{(T)},\Psi_0]
 +S_\Omega[\Omega_F^{(1)}]+S_I[I_1],
 \label{eq:V_GS_linear}
\end{equation}
where $\mathcal L_{\rm K}$ is the Kerr split-monopole operator. After
the metric perturbation has been specified, the first term is known,
but $I_1(\Psi_0)$ and $\Omega_F^{(1)}(\Psi_0)$ are fixed only by a
global solution that crosses both light surfaces.  We do not solve
that global eigenvalue problem in this work.

The boundary conditions are
\begin{align}
 &\Psi_1=0 &&\text{on the rotation axis},
 \label{eq:V_axis_bc}\\
 &\Psi_1\ \text{finite} &&\text{on the future horizon},
 \label{eq:V_horizon_bc}\\
 &\Psi_1=\order(r^{-1}) &&\text{as }r\to\infty,
 \label{eq:V_infinity_bc}
\end{align}
together, for a global solution, with regular crossing of both light
surfaces. A fixed
hemispheric-flux ensemble also imposes
$\Psi_1(\pi/2)-\Psi_1(0)=0$. This fixes the overall magnetic-flux
normalization but does not forbid angular redistribution on the
horizon.

For prescribed $I_1$ and $\Omega_F^{(1)}$, the formal Green-function
representation is
\begin{equation}
 \Psi_1(x)=\int \dd^2x'\,
 G_{\rm GS}(x,x')
 \{S_T+S_\Omega+S_I\}(x').
 \label{eq:V_GS_Green}
\end{equation}
The light-surface compatibility conditions remain part of the
definition of $G_{\rm GS}$ and of the allowed homogeneous data.  Thus
eq.~\eqref{eq:V_GS_Green} is a formal solution map, not an explicit
global EC magnetosphere.  In particular, the compact proxy source
used in an earlier draft is not used to determine any coefficient in
the present paper.  The rotating part of the global response remains
formal, whereas the static $\order(\epsilon_T\xi)$ flux redistribution
needed for the leading power coefficient is explicitly fixed by
eq.~\eqref{eq:V_static_flux_Green}.

The horizon condition follows by transforming eq.~\eqref{eq:V_F_decomposition}
to ingoing Kerr coordinates and requiring every tetrad component of
$F$ to be regular. Near the horizon, the Boyer--Lindquist one-forms
obey
\begin{equation}
 \dd t=\dd v-\frac{r^2+a^2}{\Delta}\dd r,
 \qquad
 \dd\phi=\dd\tilde\phi-\frac{a}{\Delta}\dd r.
 \label{eq:V_ingoing_forms}
\end{equation}
The apparent $\Delta^{-1}$ term generated by the first part of
eq.~\eqref{eq:V_F_decomposition} must cancel the corresponding term
in $F_{r\theta}$. Rewriting the result in terms of the invariant
current in eq.~\eqref{eq:V_current_definition} gives the structural
Znajek relation
\begin{equation}
 I_H={\cal Z}_H(\theta)
 (\Omega_H-\Omega_F)\,\partial_\theta\Psi_H,
 \label{eq:V_Znajek}
\end{equation}
where ${\cal Z}_H$ is a positive horizon geometric weight for the
orientation chosen here.  In the current convention
$I=2\pi\sqrt{-g}F^{\theta r}$, its relation to the weight used in the
power functional is, at the order retained,
\begin{equation}
 {\cal Z}_H(\theta)=2\pi{\cal W}_H(\theta),
 \qquad
 {\cal W}_H(\theta)=
 \left.\sqrt{\frac{g_{\phi\phi}}{g_{\theta\theta}}}\right|_{\cal H}
 +\order(\chi^2).
 \label{eq:V_ZW_relation}
\end{equation}
Thus the factor $2\pi$ is fixed by the definition of $I$, and the
same metric ratio controls both Znajek regularity and the horizon power
weight. Torsion does not add an independent term to
this equation: its effect enters through the metric dependence of
$\mathcal Z_H$, the corrected $\Omega_H$, and the magnetospheric
response $\Psi_H$.

Varying eq.~\eqref{eq:V_Znajek} gives
\begin{align}
 \frac{\delta I_H}{I_H^{\rm K}}=&
 \frac{\delta{\cal Z}_H}{{\cal Z}_H^{\rm K}}
 +\frac{\delta\Omega_H-\delta\Omega_F}
 {\Omega_H^{\rm K}-\Omega_F^{\rm K}}
 +\frac{\partial_\theta\delta\Psi_H}
 {\partial_\theta\Psi_H^{\rm K}}.
 \label{eq:V_Znajek_variation}
\end{align}
This identity motivates the unambiguous separation
\begin{equation}
 \delta I_H=\delta I_\Omega+\delta I_{\rm geom}+\delta I_\Psi.
 \label{eq:V_deltaI_decomp}
\end{equation}
At fixed load ratio and fixed \emph{hemispheric} flux, the
$\ell=2$ redistribution in eq.~\eqref{eq:V_static_flux_ansatz} remains
and contributes at the same relative order as the linear-$\xi$
angular-velocity term.  The explicitly projected rotation piece is
\begin{equation}
 \left.\frac{\delta I_\Omega}{I_H^{\rm K}}\right|_{x,\Phi_H}
 =\epsilon_T[C_\Omega^{(0)}+\xi C_\Omega^{(1)}]+
 \order(\epsilon_T\chi^2,\epsilon_T\xi^2),
 \label{eq:V_IOmega_controlled}
\end{equation}
where both coefficients are given by
eq.~\eqref{eq:IV_COmega_moment}.  The geometric and flux-shape terms
must be added separately.  The static flux-shape term is fixed by
eq.~\eqref{eq:V_R2H}; only the genuinely rotating global response
still requires the two-light-surface solution.

The horizon condition must finally be matched to the outer load; this
step fixes the otherwise free relation between current and field-line
rotation.

At infinity, regularity selects the Michel relation
\begin{equation}
 I=2\pi\Omega_F\left(2\Psi-\frac{\Psi^2}{\Psi_\star}\right).
 \label{eq:V_Michel}
\end{equation}
The Michel relation motivates the standard optimal-load prescription
$\Omega_F=\Omega_H/2$.  Because the perturbed global eigenvalue
problem has not been solved, we impose this relation as a load
ensemble rather than deriving its first-order persistence.  More
generally, write
\begin{equation}
 x\equiv\frac{\Omega_F}{\Omega_H},
 \qquad 0<x<1,
 \label{eq:V_load_ratio}
\end{equation}
until the power is varied. In a complete global solution the two
light-surface regularity conditions would determine $I_1$ and the
allowed change of $x$. At the imposed matched point $x=1/2$, the load factor
$x(1-x)$ is stationary. This fact is responsible for the absence of
a first-order load contribution to the leading optimal BZ power. It
does not imply that the full magnetospheric response vanishes.

\section{Analytic correction to Blandford--Znajek power}
\label{sec:BZ_power}

The horizon-flux definition and split-monopole normalization used here
are those of the analytic BZ literature
\cite{BlandfordZnajek1977,MacdonaldThorne1982,
ThornePriceMacdonald1986,GrallaJacobson2014}.

The outward electromagnetic power is the negative Killing-energy
flux through the future horizon,
\begin{equation}
 P_{\rm BZ}=-\int_{\cal H}T^r{}_t\sqrt{-g}\,\dd\theta\dd\phi.
 \label{eq:VI_power_definition}
\end{equation}
The sign convention is such that $P_{\rm BZ}>0$ denotes energy
reaching infinity. Stationarity and axisymmetry reduce the horizon
flux to a functional of the three surface quantities
$\Omega_H$, $\Omega_F$ and $\Psi_H$. Using
eq.~\eqref{eq:V_Znajek}, one obtains
\begin{equation}
 P_{\rm BZ}=2\pi\int_0^\pi\dd\theta\,
 {\cal W}_H(\theta)\Omega_F(\Omega_H-\Omega_F)
 (\partial_\theta\Psi_H)^2.
 \label{eq:VI_power_general}
\end{equation}
Here $\mathcal W_H>0$ is the same weight defined explicitly in
eq.~\eqref{eq:V_ZW_relation}; it collects the regular area and redshift
factors left after the ingoing-horizon limit. Positivity of
$\mathcal W_H(\partial_\theta\Psi_H)^2$ shows directly that outward
extraction requires $0<\Omega_F<\Omega_H$.

It is useful to isolate the angular integral
\begin{equation}
 {\cal N}_H[\Psi_H,{\cal W}_H]
 \equiv\int_0^\pi\dd\theta\,
 {\cal W}_H(\partial_\theta\Psi_H)^2.
 \label{eq:VI_NH_definition}
\end{equation}
If $x=\Omega_F/\Omega_H$ is constant on the split-monopole flux
surfaces at the order considered, eq.~\eqref{eq:VI_power_general}
becomes
\begin{equation}
 P_{\rm BZ}=2\pi\Omega_H^2x(1-x){\cal N}_H.
 \label{eq:VI_power_factorized}
\end{equation}
This factorization separates the horizon rotation, the external load,
and the magnetic and geometric angular functional.

Expanding the horizon functional gives a decomposition into geometric,
magnetic-flux, and load responses.

Write
\begin{equation}
 P_{\rm BZ}^{\rm EC}=P_{\rm BZ}^{\rm K}
 [1+\epsilon_TC_{\rm BZ}]+\order(\epsilon_T^2).
 \label{eq:VI_master_response}
\end{equation}
Define
\begin{equation}
 \frac{\delta\Omega_H}{\Omega_H}=\epsilon_TC_\Omega,
 \quad\delta x=\epsilon_TC_x,
 \label{eq:VI_response_defs}
\end{equation}
where every coefficient is evaluated on the Kerr background. For the
angular functional, let
\begin{align}
 C_{\rm hor}\equiv&\frac{1}{\epsilon_T{\cal N}_H^{\rm K}}
 \int_0^\pi\dd\theta\,
 \delta{\cal W}_H(\partial_\theta\Psi_0)^2,
 \label{eq:VI_Chor_weighted}\\
 2C_\Phi\equiv&\frac{2}{\epsilon_T{\cal N}_H^{\rm K}}
 \int_0^\pi\dd\theta\,
 {\cal W}_H^{\rm K}(\partial_\theta\Psi_0)
 (\partial_\theta\delta\Psi_H).
 \label{eq:VI_CPhi_weighted}
\end{align}
These are weighted, integrated responses. If
$\delta\Psi_H=C_\Phi\epsilon_T\Psi_0$ is a pure flux
renormalization, eq.~\eqref{eq:VI_CPhi_weighted} reduces to the usual
fractional magnetic-flux correction. A fixed hemispheric flux removes
that normalization mode, but an $\ell=2$ redistribution can still
contribute to the weighted integral. In the strict fixed-profile
approximation, rather than merely fixed total flux, one sets
$\delta\Psi_H=0$ and hence $C_\Phi=0$.

Taking the logarithmic variation of
eq.~\eqref{eq:VI_power_factorized} gives the decomposition
\begin{equation}
 C_{\rm BZ}=2C_\Omega+2C_\Phi+C_{\rm hor}+C_{\rm load},
 \label{eq:VI_CBZ_decomp}
\end{equation}
with
\begin{equation}
 C_{\rm load}=\frac{1-2x}{x(1-x)}C_x.
 \label{eq:VI_Cload}
\end{equation}
Thus $C_{\rm load}=0$ at $x=1/2$ to first order even when
$C_x\neq0$. In the fixed-profile approximation, $C_\Phi=0$, and the
purely geometric part reduces to
$C_{\rm BZ}^{\rm geom}=2C_\Omega+C_{\rm hor}$. No assumption about
the sign of these coefficients has entered the derivation.

For the split-monopole seed the angular integral can be evaluated
exactly and the slow-rotation order of every term becomes explicit.

At leading slow-rotation order,
\begin{equation}
 \Omega_H^{\rm K}=\frac{\chi}{4M}+\order(\chi^3),
 \qquad
 {\cal W}_H^{\rm K}=\sin\theta+\order(\chi^2).
 \label{eq:VI_Kerr_leading_data}
\end{equation}
For a circular horizon, eq.~\eqref{eq:V_ZW_relation} makes the direct
horizon-weight response transparent.  Equation~\eqref{eq:V_static_polar_metric} has
$h_{\phi\phi}^{(2)}=\sin^2\theta\,h_{\theta\theta}^{(2)}$; evaluating
both components on the shifted horizon therefore leaves their ratio
equal to $\sin^2\theta$ through $\order(\epsilon_T\xi)$.  Hence
\begin{equation}
 C_{\rm hor}^{(1)}=0
 \label{eq:VI_Chor1_zero}
\end{equation}
in this fixed Regge--Wheeler/circular gauge, while the integrated
physical response remains gauge invariant.
Because $\partial_\theta\Psi_0=\Psi_\star\sin\theta$, the angular
functional is
\begin{equation}
 {\cal N}_H^{\rm K}
 =\Psi_\star^2\int_0^\pi\sin^3\theta\,\dd\theta
 =\frac{4}{3}\Psi_\star^2+\order(\chi^2).
 \label{eq:VI_NK_split}
\end{equation}
The magnetostatic solution gives on the horizon
\begin{equation}
 \delta\Psi_H^{(\xi)}
 =\epsilon_T\xi\Psi_\star R_{2H}X_\phi^{20}.
 \label{eq:VI_deltaPsi_static}
\end{equation}
Using
\begin{equation}
 \int_0^\pi\!d\theta\,\sin^2\theta\,
 \partial_\theta X_\phi^{20}=-\frac85,
 \label{eq:VI_flux_angular_integral}
\end{equation}
in eq.~\eqref{eq:VI_CPhi_weighted} yields
\begin{equation}
 C_\Phi=\xi C_\Phi^{(1)}+\order(\xi^2),
 \qquad C_\Phi^{(1)}=-\frac65R_{2H}.
 \label{eq:VI_CPhi1}
\end{equation}
Thus fixed total flux does not set $C_\Phi^{(1)}$ to zero.
Equation~\eqref{eq:VI_power_factorized} therefore gives
\begin{equation}
 P_{\rm BZ}^{\rm K}
 =\frac{8\pi}{3}\Psi_\star^2\Omega_H^2x(1-x)
 +\order(\chi^4).
 \label{eq:VI_PKerr_general_load}
\end{equation}
The factor $x(1-x)$ reaches its maximum at $x=1/2$. Substituting that
value gives, in our split-monopole normalization,
\begin{equation}
 P_{\rm BZ}^{\rm K}=\frac{\pi\Psi_\star^2}{24M^2}\chi^2
 +\order(\chi^4).
 \label{eq:VI_PKerr}
\end{equation}
The EC response now has a closed leading hierarchy.  The explicitly
completed rotational source gives $C_\Omega^{(0)}+\xi C_\Omega^{(1)}$
in eq.~\eqref{eq:IV_COmega_moment}.  The direct horizon-weight term
vanishes by eq.~\eqref{eq:VI_Chor1_zero}, whereas the static
$\ell=2$ magnetic-flux redistribution contributes through
eq.~\eqref{eq:VI_CPhi1}.  At optimal load the linear load response
also vanishes. Consequently, at absolute order
$\order(\epsilon_T\chi^2)$,
\begin{equation}
 C_{\rm BZ}^{(0)}+\xi C_{\rm BZ}^{(1)}
 =2[C_\Omega^{(0)}+\xi C_\Omega^{(1)}]
 +2\xi C_\Phi^{(1)},
 \label{eq:VI_leading_origin}
\end{equation}
The central controlled result for the stated source, fixed-ADM
subtraction, fixed flux and imposed optimal load is therefore
\begin{equation}
 {
 C_{\rm BZ}^{(0)}+\xi C_{\rm BZ}^{(1)}
 =\frac{32\pi\lambda_J}{3}
 {\cal I}_J(q,w)
 +\xi\left[-\frac{64\pi\lambda_J}{15}{\cal I}_J(q,w)
 -\frac{12}{5}R_{2H}(q,w)\right]}.
 \label{eq:VI_CBZ0}
\end{equation}
There is no universal positive coefficient: the sign and magnitude
are inherited from both the specified rotational completion and the
static polar response.  The first linear-$\xi$ term is fixed by the
axial $\ell=1$ projection in
eq.~\eqref{eq:IV_axial_decomposition}; the second is the
magnetostatic redistribution in eq.~\eqref{eq:V_R2H}.  The
simultaneously generated axial $\ell=3$ mode vanishes at the horizon.
The omitted
$\order(\epsilon_T\chi^4)$ power requires both the second-order
horizon geometry and the global light-surface-matched flux response;
we do not quote a partial coefficient for that order.

Figure~\ref{fig:bz-power} illustrates the leading model-dependent response
in the $(\lambda_J,\xi)$ plane for $q=0$ and $w=0.8$. The contour
$C_{\rm BZ}=0$ separates enhancement from suppression of the
Blandford--Znajek power. The figure is a direct evaluation of
eq.~\eqref{eq:VI_CBZ0}, not an independent numerical fit.

\begin{figure}[t]
    \centering
    \includegraphics[width=0.82\textwidth]{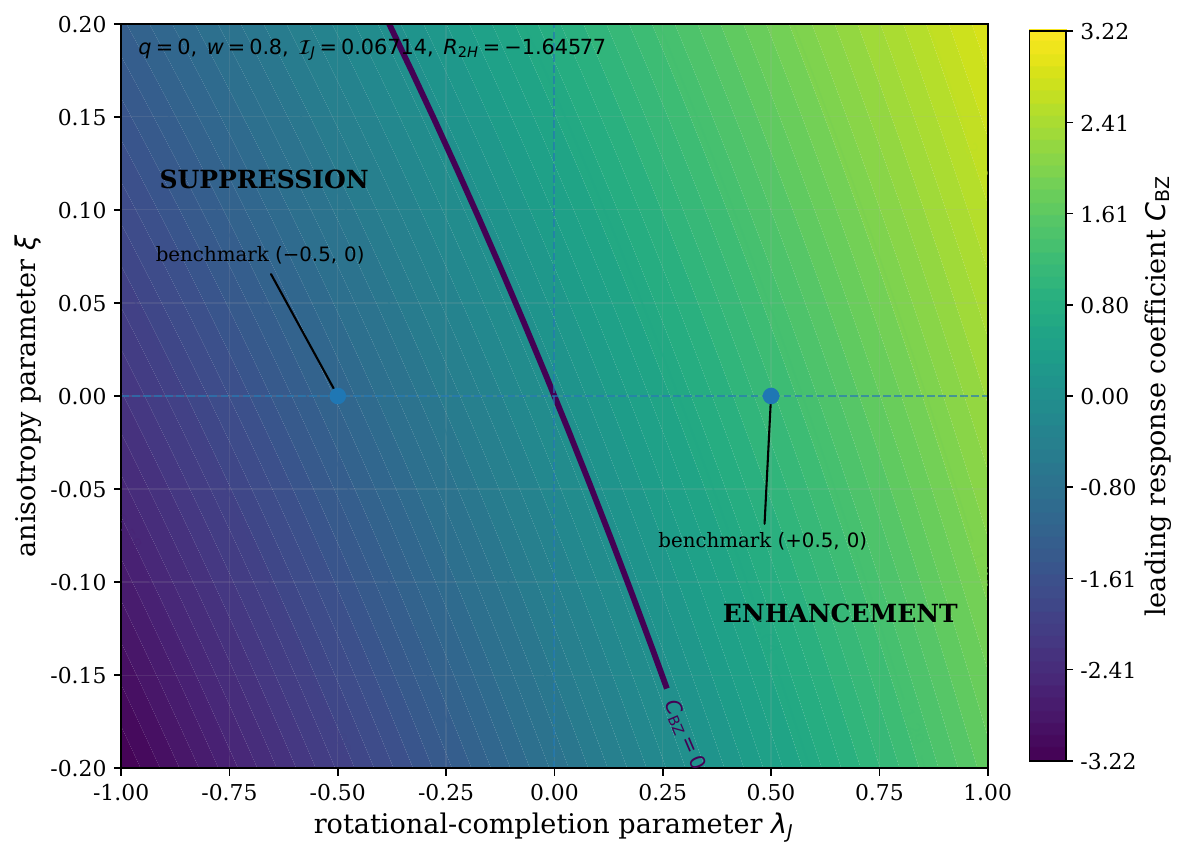}
    \caption{
    Leading Blandford--Znajek response in the $(\lambda_J,\xi)$ plane
    for the representative source parameters $q=0$ and $w=0.8$.
    The color scale shows the coefficient $C_{\rm BZ}$ of eq.~(9.2),
    evaluated with $\mathcal{I}_J=0.06714$ and $R_{2H}=-1.64577$.
    The contour $C_{\rm BZ}=0$ separates regions of enhanced
    ($C_{\rm BZ}>0$) and suppressed ($C_{\rm BZ}<0$)
    Blandford--Znajek power. The marked points
    $(\lambda_J,\xi)=(\pm0.5,0)$ indicate the representative isotropic
    co-rotating and counter-rotating completions.
    }
    \label{fig:bz-response-phase}
\end{figure}

\section{Analytic properties and limiting cases}
\label{sec:analytic_limits}

The limits below combine the standard slow-rotation onset of
split-monopole BZ power with the algebraic, nonpropagating character
of minimally coupled EC torsion
\cite{BlandfordZnajek1977,TchekhovskoyNarayanMcKinney2010,
Hehl1976,Shapiro2002}.

Equation~\eqref{eq:VI_master_response} immediately gives the GR
limit
\begin{equation}
 \epsilon_T\to0\quad\Longrightarrow\quad
 P_{\rm BZ}^{\rm EC}\to P_{\rm BZ}^{\rm K}.
 \label{eq:VII_GR_limit}
\end{equation}
Likewise the exact Cartan relation and the normalization of the
effective source imply
\begin{equation}
 J_5^\mu\to0\quad\Longrightarrow\quad
 K_{\alpha\beta\gamma},\ T^\lambda{}_{\mu\nu},\
 \tau_{\mu\nu}^{(T)},\ h_{\mu\nu}^{(T)}\to0.
 \label{eq:VII_spinless}
\end{equation}
Finally,
\begin{equation}
 \chi\to0\quad\Longrightarrow\quad P_{\rm BZ}\to0,
 \label{eq:VII_nonrotating}
\end{equation}
because $\Omega_H=\order(\chi)$ and the BZ power begins at
$\order(\chi^2)$. The fractional response is understood as the limit
from nonzero spin.

The order counting is summarized in table~\ref{tab:order_counting}.
The complete term linear in $\xi$ in the leading power response is
controlled here, including its static magnetostatic part; coefficients at relative
$\order(\epsilon_T\chi^2)$ require the global magnetosphere.

\begin{table}[t]
 \centering
 \begin{tabular}{c|c|c}
  quantity & Kerr scaling & controlled EC information \\
  \hline
  $\Omega_H$ & $\order(\chi)$ &
  $\delta\Omega_H/\Omega_H=
  \epsilon_T(C_\Omega^{(0)}+\xi C_\Omega^{(1)})$ \\
  ${\cal W}_H$ & $\order(1)$ &
  $C_{\rm hor}^{(1)}=0$ in the circular gauge \\
  $\Psi_H$ & $\order(1)$ &
  $\delta\Psi_H=\epsilon_T\xi\Psi_\star R_{2H}X_\phi^{20}$ \\
  $x(1-x)$ at $x=1/2$ & $\order(1)$ &
  first variation $0$ \\
  $P_{\rm BZ}$ & $\order(\chi^2)$ &
  $\delta P_{\rm BZ}=\order(\epsilon_T\chi^2)$
 \end{tabular}
 \caption{Consistent slow-rotation and torsion order counting for the
 imposed optimal-load ensemble.}
 \label{tab:order_counting}
\end{table}

To avoid mixing absolute and relative orders, write
\begin{align}
 P_{\rm BZ}^{\rm EC}=&A_2\chi^2
 [1+\epsilon_TC_0(q,w,\xi,\lambda_J)]
 \nonumber\\
 &+A_4\chi^4[1+\epsilon_TC_2]+\cdots.
 \label{eq:VII_slow_expansion}
\end{align}
Equation~\eqref{eq:VI_leading_origin} gives
\begin{equation}
 {C_0(q,w,\xi,\lambda_J)=
 \frac{32\pi\lambda_J}{3}{\cal I}_J(q,w)
 +\xi\left[-\frac{64\pi\lambda_J}{15}{\cal I}_J(q,w)
 -\frac{12}{5}R_{2H}(q,w)\right]}.
 \label{eq:VII_C0_model}
\end{equation}
The rotational part of the linear-$\xi$ coefficient follows from the
axial $\ell=1$ projection in eq.~\eqref{eq:IV_axial_decomposition};
the independent $R_{2H}$ term follows from the static polar metric.
Neither term can be inferred from the scalar monopole average of
$P_2$.
Thus a leading-order no-correction theorem is not implied by the EC
field equations alone.  At $\xi=0$, $C_0$ vanishes when
$\lambda_J=0$; for $\xi\ne0$, a static magnetostatic contribution can
remain even without a rotational supporting current. This conditional
statement replaces the source-independent positive coefficient
asserted in the previous version.

The radial moment has two useful analytic properties. For $w>0$,
\begin{equation}
 {\cal I}_J(q,w)>0,
 \qquad
 \frac{\partial {\cal I}_J}{\partial q}
 =2\int_2^{2+w}\!\dd y\,\ln y\,
 y^{2q}{\cal W}_{\rm NH}^2(y)
 \left(\frac{y^2}{8}-\frac1y\right)>0.
 \label{eq:VII_IJ_properties}
\end{equation}
The second inequality describes the radial weighting only; the sign
of the physical response also contains $\lambda_J$.

Because $\epsilon_T=\kappa^2J_0^2M^2\geq0$ in the present
parameterization, the exact leading sign criterion for this closure is
\begin{equation}
 \frac{32\pi\lambda_J}{3}{\cal I}_J
 +\xi\left[-\frac{64\pi\lambda_J}{15}{\cal I}_J
 -\frac{12}{5}R_{2H}\right]
 \begin{cases}>0,&\text{enhancement},\\<0,&\text{suppression}.
 \end{cases}
 \label{eq:VII_sign_model}
\end{equation}
For a general completion, $C_{\rm BZ}>0$ gives enhancement and
$C_{\rm BZ}<0$ gives suppression. A convenient sufficient condition
at the order where all four response terms are known is
\begin{equation}
 C_\Omega<0,\qquad
 |2C_\Omega|>|2C_\Phi+C_{\rm hor}+C_{\rm load}|.
 \label{eq:VII_suppression_sufficient}
\end{equation}
We do not extrapolate this inequality to a high-spin phase diagram.

For a chosen tolerance $0<\eta\ll1$, perturbative control requires
\begin{equation}
 \left|\epsilon_T\left\{\frac{32\pi\lambda_J}{3}{\cal I}_J
 +\xi\left[-\frac{64\pi\lambda_J}{15}{\cal I}_J
 -\frac{12}{5}R_{2H}\right]\right\}\right|<\eta,
 \qquad |\xi|\ll1,\qquad \chi^2\ll1.
 \label{eq:VII_validity_bound}
\end{equation}
Equation~\eqref{eq:VII_C0_model} contains the complete linear
anisotropy correction of the stated axial and polar source.  It is not an exact
$\xi\to-\xi$ symmetry, and the quadratic dependence still requires
the complete $\ell=0,2,4$ polar and $\ell=1,3,5$ axial systems at
$\order(\xi^2)$.

\section{First-law bookkeeping and energy consistency}
\label{sec:first_law}

The distinction between a first-law identity and an independent
matter-energy computation follows the covariant phase-space and
Noether-charge framework
\cite{BardeenCarterHawking1973,Wald1993,IyerWald1994,
SudarskyWald1992,WaldZoupas2000}.

For a stationary perturbation with a localized material source, the
horizon first law is
\begin{equation}
 \delta M=\frac{\kappa_H}{8\pi G}\delta A_H
 +\Omega_H\delta J+\delta E_{\rm spin}.
 \label{eq:VIII_first_law}
\end{equation}
Our comparison branch sets
$\delta M_{\rm ADM}=\delta J_{\rm ADM}=0$ by subtracting the
homogeneous Kerr parameter modes described in
section~\ref{sec:near_horizon_geometry}.  The first law then
\emph{defines} the total source-plus-matching energy required by this
ensemble,
\begin{equation}
 {\delta E_{\rm req}
 \equiv-\frac{\kappa_H^{\rm K}}{8\pi G}\delta A_H}.
 \label{eq:VIII_Erequired}
\end{equation}
This is bookkeeping, not an independent verification of the first
law.  The effective stress tensor determines the metric and hence
$\delta A_H$, but it does not provide a microscopic Hamiltonian for
the compensating matching system.  An independent check would require
computing the matter and matching contributions from a global Dirac
solution and its covariant phase-space or Noether charge.  We make no
such claim here.

The physical meaning of the fixed-ADM branch is correspondingly
limited.  It compares geometries with the same charges measured at
infinity after the source mass and angular momentum have been
compensated by the homogeneous parameter subtraction.  It may be
realized by an external support or matching layer, but the present
near-horizon model does not choose between those microscopic
realizations.  Results on this branch are therefore ensemble- and
source-model dependent.

Stationarity and axisymmetry imply that energy and angular-momentum
fluxes are locked on each magnetic surface:
\begin{equation}
 \dd P_{\rm BZ}=\Omega_F\,\dd\dot J_{\rm EM}.
 \label{eq:VIII_flux_identity}
\end{equation}
For the Michel asymptotic solution, the Mathematica tensor
contraction gives
\begin{align}
 {\cal E}^r&=\frac{\Omega_F^2\Psi_\star^2\sin^2\theta}
 {4\pi r^2},
 \label{eq:VIII_Eflux}\\
 {\cal J}^r&=\frac{\Omega_F\Psi_\star^2\sin^2\theta}
 {4\pi r^2},
 \label{eq:VIII_Jflux}
\end{align}
so ${\cal E}^r-\Omega_F{\cal J}^r=0$ identically, rather than only
after angular integration.

The flux identity must be supplemented by the sign condition for
outward energy extraction.

Let $\chi_H^\mu=t^\mu+\Omega_H\phi^\mu$. Negative Killing-energy
flux into the horizon is
\begin{equation}
 T_{\mu\nu}^{\rm EM}t^\mu\chi_H^\nu\big|_{\cal H}<0,
 \label{eq:VIII_negative_energy}
\end{equation}
which is equivalent for the regular force-free solution to
\begin{equation}
 0<\Omega_F<\Omega_H.
 \label{eq:VIII_extraction_range}
\end{equation}
At optimal load $\Omega_F=\Omega_H/2$ and therefore
\begin{equation}
 \Omega_F(\Omega_F-\Omega_H)=-\frac{\Omega_H^2}{4}<0.
 \label{eq:VIII_extraction_product}
\end{equation}
For either sign of $\lambda_J$, the corrected $\Omega_H$ keeps the
sign of its Kerr value whenever
$|\epsilon_T(C_\Omega^{(0)}+\xi C_\Omega^{(1)})|\ll1$.
If the imposed load satisfies
$0<x<1$, then $0<\Omega_F<\Omega_H$ and the extraction interval is
preserved at the controlled order.  This electromagnetic statement
is independent of the first-law bookkeeping above.

\section{Discussion and conclusions}
\label{sec:conclusion}

Our interpretation retains the standard EC distinction between local,
algebraic torsion in matter and an exterior metric response
\cite{Hehl1976,Shapiro2002,Trautman2006,LuzMena2025,
KatkarPhadatare2026}, together with the usual BZ
requirement of a globally matched force-free magnetosphere
\cite{BlandfordZnajek1977,Komissarov2004,GrallaJacobson2014,
MeringoloCamilloniRezzolla2025,FengCaiYangZhang2026}.

In minimal EC theory the axial Dirac current produces a local,
totally antisymmetric contortion and a four-fermion contact term;
there is no propagating torsion field in vacuum. For an inhomogeneous
profile that contact stress is not independently conserved. The
central modeling step of this work is therefore an explicitly stated,
compact, conserved phenomenological completion. It is inspired by the
EC--Dirac system but is not presented as a global solution of the
Hehl--Datta equation.  The torsion-producing species is neutral,
$q_{\rm spin}=0$, so it contributes no Dirac vector current to the
Maxwell equation; the BZ current is carried by the separate force-free
plasma.

The correct scientific statement is consequently narrower than a
universal torsion correction: for the specified conserved
phenomenological completion, a rotational matter current at
$\order(\epsilon_T\chi)$ produces the conditional leading BZ response
below.  Equation~\eqref{eq:III_lambdaJ_interpretation} gives a
coarse-grained Weyssenhoff-fluid interpretation of $\lambda_J$, but
its numerical value still requires an equation of state, spin
transport and a matter rotation profile
\cite{Hehl1976,HehlDatta1971,ObukhovKorotky1987,
AgarwalCarloni2025,VignoloDeMariaFabbriCarloni2025,
AbboudGavassinoSinghSperanza2025}.

The response can be summarized as
\begin{equation}
 {P_{\rm BZ}^{\rm EC}=P_{\rm BZ}^{\rm K}
 \left[1+\epsilon_T
 C_{\rm BZ}[\tau_{\mu\nu}^{(T)},x]\right]}
 \label{eq:IX_final}
\end{equation}
At fixed ADM charges, fixed hemispheric flux and imposed optimal load
$x=\Omega_F/\Omega_H=1/2$, the controlled leading slow-rotation
coefficient for the specified completion is
\begin{equation}
 C_{\rm BZ}=\frac{32\pi\lambda_J}{3}
 {\cal I}_J(q,w)
 +\xi\left[-\frac{64\pi\lambda_J}{15}{\cal I}_J(q,w)
 -\frac{12}{5}R_{2H}(q,w)\right]
 +\order(\chi^2,\xi^2).
 \label{eq:IX_final_coefficient}
\end{equation}
At $\xi=0$ its sign is the sign of $\lambda_J$; at nonzero $\xi$ the
static flux functional $R_{2H}$ also enters, so EC alone does not
select universal enhancement or suppression. The strict
$\order(\xi)$ expansion now includes both the axial $\ell=1$
projection and the polar magnetostatic response. We neither use a
partial resummation in $\chi$ nor draw a
high-spin phase diagram.

The effective domain is
\begin{equation}
 |\epsilon_T|\ll1,\qquad |\epsilon_TC_i|\ll1,
 \qquad \chi^2\ll1,\qquad w>0,
 \label{eq:IX_validity}
\end{equation}
with the extremal limit excluded from the uniform expansion. The
generalized Znajek relation is local and controlled, whereas the
linearized Grad--Shafranov Green representation remains formal until
the two light surfaces are matched. Consequently the present result
is a leading slow-rotation response for a stated source and load ensemble,
not a complete global EC force-free solution. The first law fixes the
total energy required by the fixed-ADM comparison but does not
independently compute it.

\appendix

\section{Technical properties of the near-horizon spin source}
\label{app:spin_source_checks}

This appendix supplies component-level checks of the EC source model;
the geometric conventions are those of refs.~\cite{Hehl1976,
Shapiro2002}.
For completeness, we collect here several identities used in the
construction of the stationary and axisymmetric spin source of
section~\ref{sec:spin_source}.

The radial window introduced in eq.~\eqref{eq:III_window} can be
written as
\begin{equation}
    {\cal W}_{\rm NH}(x)
    =
    \begin{cases}
    \displaystyle
    \exp\left(
        1-\frac{1}{1-x^{2}}
    \right),
    &
    0\leq x<1,
    \\[1.5ex]
    0,
    &
    x\geq1.
    \end{cases}
    \label{eq:app_window}
\end{equation}
At the horizon,
\begin{equation}
    {\cal W}_{\rm NH}(0)=1,
    \qquad
    \left.
    \frac{d{\cal W}_{\rm NH}}{dx}
    \right|_{x=0}
    =
    0.
\end{equation}
At the outer boundary,
\begin{equation}
    \lim_{x\rightarrow1^{-}}
    {\cal W}_{\rm NH}(x)
    =
    0,
\end{equation}
and, more generally,
\begin{equation}
    \lim_{x\rightarrow1^{-}}
    \frac{d^{n}{\cal W}_{\rm NH}}
         {dx^{n}}
    =
    0,
    \qquad
    n=1,2,\ldots .
    \label{eq:app_window_smoothness}
\end{equation}
The source therefore possesses a $C^\infty$ matching to the exterior
region.

For symbolic checks it is useful to choose a simple local
polarization vector.  In Boyer--Lindquist coordinates we used
\begin{equation}
    s_{\rm pol}^{\mu}
    =
    \left(
        0,
        0,
        \frac{1}{\sqrt{\Sigma}},
        0
    \right).
    \label{eq:app_spolar}
\end{equation}
It obeys
\begin{equation}
    \bar g_{\mu\nu}
    s_{\rm pol}^{\mu}s_{\rm pol}^{\nu}
    =
    1.
\end{equation}
For the circular four-velocity
eq.~\eqref{eq:III_circular_velocity},
\begin{equation}
    u_{\mu}s_{\rm pol}^{\mu}=0
\end{equation}
follows immediately from the block-diagonal structure of the
$r$--$\theta$ sector of the Kerr metric.  This representative is used
only for explicit component checks.  The scalar source
$J_5^2={\cal A}^2$, and hence
$U_{\mu\nu}^{\rm spin}$, is independent of this particular
polarization basis.  A globally defined physical spin configuration
may instead be specified in a regular orthonormal tetrad.

With the orientation
\begin{equation}
    \epsilon_{tr\theta\phi}
    =
    +\sqrt{-\bar g}
    =
    \Sigma\sin\theta,
    \label{eq:app_orientation}
\end{equation}
the polar representative has a single independent class of
contortion components.  In particular,
\begin{align}
    K_{tr\phi}
    &=
    -\frac{\kappa}{4}
    \epsilon_{tr\phi\theta}
    J_{5}^{\theta}
    \nonumber\\
    &=
    \frac{\kappa J_{0}}{4}
    \left(
        \frac{r}{M}
    \right)^{q}
    {\cal W}_{\rm NH}(r)
    f(\theta)
    \sqrt{\Sigma}\sin\theta.
    \label{eq:app_Ktrphi}
\end{align}
The remaining nonzero components follow from complete
antisymmetry,
\begin{equation}
    K_{\alpha\beta\gamma}
    =
    K_{[\alpha\beta\gamma]}.
\end{equation}
Equation~\eqref{eq:app_Ktrphi} vanishes at the outer edge of the
source together with all derivatives of the radial window.

The multipolar identity used in
Equation~\eqref{eq:III_multipoles_exact} follows from
\begin{equation}
    P_{2}^{2}(\mu)
    =
    \frac{1}{5}P_{0}(\mu)
    +
    \frac{2}{7}P_{2}(\mu)
    +
    \frac{18}{35}P_{4}(\mu).
    \label{eq:app_P2sq}
\end{equation}
Therefore
\begin{align}
    \left(
        1+\xi P_{2}
    \right)^{2}
    =&
    P_{0}
    +
    2\xi P_{2}
    +
    \xi^{2}P_{2}^{2}
    \nonumber\\
    =&
    \left(
        1+\frac{\xi^{2}}{5}
    \right)P_{0}
    +
    \left(
        2\xi+\frac{2\xi^{2}}{7}
    \right)P_{2}
    +
    \frac{18\xi^{2}}{35}P_{4}.
    \label{eq:app_multipole_derivation}
\end{align}
This also shows explicitly that the $\ell=4$ sector begins only at
${\cal O}(\xi^{2})$.

It is also useful to display the logarithmic gradients of the scalar
source.  Inside its support,
\begin{equation}
    {\cal S}
    \propto
    r^{2q}
    {\cal W}_{\rm NH}^{2}(r)
    f^{2}(\theta),
\end{equation}
and hence
\begin{equation}
    \frac{\partial_r{\cal S}}{{\cal S}}
    =
    \frac{2q}{r}
    +
    2
    \frac{{\cal W}_{\rm NH}'}
         {{\cal W}_{\rm NH}},
    \label{eq:app_radial_gradient}
\end{equation}
while
\begin{equation}
    \frac{\partial_\theta{\cal S}}{{\cal S}}
    =
    2
    \frac{f'(\theta)}
         {f(\theta)}.
    \label{eq:app_angular_gradient}
\end{equation}
For
$f=1+\xi P_2(\cos\theta)$,
\begin{equation}
    \frac{\partial_\theta{\cal S}}{{\cal S}}
    =
    -
    \frac{
        6\xi\sin\theta\cos\theta
    }{
        1+\xi P_{2}(\cos\theta)
    }.
    \label{eq:app_angular_gradient_explicit}
\end{equation}

Finally, because metric compatibility gives
$\bar\nabla_\alpha\bar g_{\mu\nu}=0$,
the divergence of the explicit contact contribution is
\begin{align}
    \bar\nabla^\mu
    U_{\mu\nu}^{\rm spin}
    &=
    \bar\nabla^\mu
    \left(
       {\cal S}\bar g_{\mu\nu}
    \right)
    \nonumber\\
    &=
    \partial_\nu{\cal S}.
    \label{eq:app_divU}
\end{align}
For a stationary axisymmetric source,
\begin{equation}
    \partial_t{\cal S}
    =
    \partial_\phi{\cal S}
    =
    0,
\end{equation}
but in general
\begin{equation}
    \partial_r{\cal S}\neq0,
    \qquad
    \partial_\theta{\cal S}\neq0.
\end{equation}
Thus $U_{\mu\nu}^{\rm spin}$ cannot by itself be used as the source
of the linearized Einstein equation for the inhomogeneous
configuration considered here.  The linearized Bianchi identity,
\begin{equation}
    \bar\nabla^\mu
    \delta G_{\mu\nu}[h]
    =
    0,
\end{equation}
requires
\begin{equation}
    \bar\nabla^\mu
    \widehat\tau_{\mu\nu}^{(T)}
    =
    0,
\end{equation}
with the normalized conserved source defined in
eq.~\eqref{eq:III_normalized_source}. This condition is the covariant statement
that the spin--spin interaction exchanges energy and momentum with
the remaining Dirac sector rather than constituting an independently
conserved fluid.

Although Boyer--Lindquist coordinates are convenient for separating
the stationary and axisymmetric equations, they are singular at
$\Delta=0$.  Horizon regularity should therefore be assessed after
the transformation to ingoing Kerr coordinates,
\begin{equation}
    dv
    =
    dt
    +
    \frac{r^{2}+a^{2}}{\Delta}\,dr,
    \qquad
    d\tilde{\phi}
    =
    d\phi
    +
    \frac{a}{\Delta}\,dr.
    \label{eq:app_ingoing}
\end{equation}
The source scalars
$J_5^2$, ${\cal S}$, and $\epsilon_T^{\rm loc}$ remain finite at
$r=r_+$, and the apparent divergence of individual
Boyer--Lindquist components such as
$U_{rr}^{\rm spin}={\cal S}\bar g_{rr}$ is therefore a coordinate
artifact rather than a physical singularity.

\section{Stationary axisymmetric metric perturbations}
\label{app:metric_perturbation}

The polar and axial decompositions use the Regge--Wheeler--Zerilli
framework and its gauge-invariant refinements
\cite{ReggeWheeler1957,Zerilli1970,Moncrief1974,
Chandrasekhar1983,MartelPoisson2005}.

This appendix collects technical details of the stationary and
axisymmetric perturbation problem used in
section~\ref{sec:near_horizon_geometry}.

A general first-order coordinate transformation
\begin{equation}
    x^\mu
    \rightarrow
    x^\mu
    -
    \epsilon_T
    \xi^\mu
    \label{eq:appIV_gauge_transform}
\end{equation}
changes the metric perturbation according to
\begin{equation}
    h_{\mu\nu}
    \rightarrow
    h_{\mu\nu}
    +
    2\bar\nabla_{(\mu}\xi_{\nu)}.
    \label{eq:appIV_h_gauge}
\end{equation}
For stationary and axisymmetric perturbations,
\begin{equation}
    \partial_t\xi^\mu
    =
    \partial_\phi\xi^\mu
    =
    0.
\end{equation}
The four components of $\xi^\mu$ may be used to eliminate redundant
metric functions.  In the circular and reflection-symmetric sector
considered in the main text we choose the gauge
\begin{equation}
    h_{tr}
    =
    h_{t\theta}
    =
    h_{r\theta}
    =
    h_{r\phi}
    =
    h_{\theta\phi}
    =
    0,
    \label{eq:appIV_gauge_choice}
\end{equation}
where the additional vanishing condition follows from circularity of
the stationary axisymmetric configuration.

For the static even-parity $\ell=2$ sector on the Schwarzschild
background, the metric perturbation can be written as
\begin{align}
    h_{tt}^{(2)}
    &=
    f(r)
    H_0^{(2)}(r)
    P_2(\cos\theta),
    \\
    h_{rr}^{(2)}
    &=
    f^{-1}(r)
    H_2^{(2)}(r)
    P_2(\cos\theta),
    \\
    h_{\theta\theta}^{(2)}
    &=
    r^2
    K^{(2)}(r)
    P_2(\cos\theta),
    \\
    h_{\phi\phi}^{(2)}
    &=
    r^2\sin^2\theta
    K^{(2)}(r)
    P_2(\cos\theta),
    \label{eq:appIV_even_metric}
\end{align}
where
\begin{equation}
    f(r)
    =
    1-\frac{2M}{r}.
\end{equation}

We define
\begin{equation}
    \lambda
    =
    \frac12
    (\ell-1)(\ell+2),
    \label{eq:appIV_lambda}
\end{equation}
so that
\begin{equation}
    \lambda=2
\end{equation}
for $\ell=2$.  A convenient Zerilli--Moncrief master variable is
\begin{equation}
    Z_{\ell}
    =
    \frac{r}{\lambda r+3M}
    \left[
        K^{(\ell)}
        +
        \frac{
            f(r)
        }{
            \lambda r+3M
        }
        \left(
            H_2^{(\ell)}
            -
            r\frac{dK^{(\ell)}}{dr}
        \right)
    \right].
    \label{eq:appIV_Zdefinition}
\end{equation}
For a static source it obeys
\begin{equation}
    \frac{d^2Z_\ell}{dr_*^2}
    -
    V_Z^{(\ell)}(r)
    Z_\ell
    =
    {\cal S}_{Z}^{(\ell)}(r),
    \label{eq:appIV_Zeq}
\end{equation}
where
\begin{align}
    V_Z^{(\ell)}
    =
    \frac{
        2f(r)
    }{
        r^3
        (\lambda r+3M)^2
    }
    \Big[
        &
        \lambda^2(\lambda+1)r^3
        +
        3\lambda^2Mr^2
        \nonumber\\
        &
        +
        9\lambda M^2r
        +
        9M^3
    \Big].
    \label{eq:appIV_Zpotential_general}
\end{align}
For $\ell=2$, $\lambda=2$, and therefore
\begin{equation}
    V_Z^{(2)}
    =
    \frac{
        2f(r)
    }{
        r^3(2r+3M)^2
    }
    \left(
        12r^3
        +
        12Mr^2
        +
        18M^2r
        +
        9M^3
    \right).
    \label{eq:appIV_Zpotential2}
\end{equation}

The effective source can be expanded in scalar spherical harmonics as
\begin{equation}
    \widehat\tau_{\mu\nu}^{(T)}
    =
    \sum_{\ell}
    \widehat\tau_{\mu\nu}^{(\ell)}(r)
    P_\ell(\cos\theta),
    \label{eq:appIV_tau_harmonics}
\end{equation}
with
\begin{equation}
    \ell=0,2
\end{equation}
in the scalar/even-parity sector at ${\cal O}(\xi)$.  To make the
source and reconstruction completely explicit, introduce
\begin{equation}
 s(y)\equiv\kappa M^2\widehat{\cal S}_0(y)
 =\frac{3}{16}y^{2q}{\cal W}_{\rm NH}^2(y),
 \qquad a(y)\equiv2s(y)+ys'(y).
 \label{eq:appIV_sa}
\end{equation}
Primes in the remainder of this polar reduction denote $d/dy$.
The coefficient of $\xi$ in the normalized mixed source is
\begin{align}
 \kappa M^2[\widehat\tau^t{}_t]_{\xi}
 &=\kappa M^2[\widehat\tau^r{}_r]_{\xi}
 =2sP_2,
 \nonumber\\
 \kappa M^2[\widehat\tau^\theta{}_\theta]_{\xi}
 &=\frac{a}{2}(1+P_2),
 &
 \kappa M^2[\widehat\tau^\phi{}_\phi]_{\xi}
 &=\frac{a}{2}(3P_2-1),
 \label{eq:appIV_explicit_polar_source}
\end{align}
with vanishing static shears.  Direct substitution gives
\begin{equation}
    \bar\nabla^\mu\widehat\tau_{\mu\nu}^{(T)}
    =\order(\xi^2,\chi^2)
    \label{eq:appIV_completion_conservation}
\end{equation}
through the order used here.

The mixed Einstein equations reduce directly to a single equation,
\begin{equation}
 {\cal L}_K[K^{(2)}]={\cal S}_K,
 \qquad
 {\cal L}_K[K]=(y-2)y(y+1)K''+(2y^2+y-4)K'-6yK,
 \label{eq:appIV_K_equation}
\end{equation}
where
\begin{equation}
 {\cal S}_K(y)=-y\left\{
 4(y^2+2y-2)s+y(8y^2+y-10)s'
 +y^2(y-2)(y+1)s''\right\}.
 \label{eq:appIV_K_source}
\end{equation}
Two real homogeneous solutions adapted to the boundaries are
\begin{align}
 v_H(y)&=y^2-2,
 \nonumber\\
 v_\infty(y)&=\frac{4-6y-6y^2
 +3y(2-y^2)\ln[(y-2)/y]}{16y}.
 \label{eq:appIV_K_homogeneous}
\end{align}
They obey $v_H(2)=2$, $v_\infty(\infty)=0$ and
\begin{equation}
 W_K\equiv v_Hv_\infty'-v_\infty v_H'
 =-\frac{y+1}{(y-2)y^2}.
 \label{eq:appIV_K_Wronskian}
\end{equation}
Consequently,
\begin{align}
 G_K(y,z)&=-\frac{z}{(z+1)^2}
 v_H(y_<)v_\infty(y_>),
 \nonumber\\
 K^{(2)}(y)&=\int_2^\infty\!dz\,G_K(y,z){\cal S}_K(z),
 \label{eq:appIV_K_Green}
\end{align}
where $y_<=\min(y,z)$ and $y_>=\max(y,z)$.  This kernel has unit
derivative jump under ${\cal L}_K$, is regular at the future horizon,
and yields the decaying exterior multipole.

The remaining functions follow without another integration:
\begin{align}
 H_2^{(2)}={}&\frac{y}{2(y+1)}
 \left[2y^2s-y(2y+3)a+2K^{(2)}-(K^{(2)})'\right],
 \nonumber\\
 H_0^{(2)}={}&H_2^{(2)}+y^2a.
 \label{eq:appIV_H_reconstruction}
\end{align}
In particular, the metric combination sourcing the static Maxwell
problem is explicitly
\begin{equation}
 {\beta_2(y)=-K^{(2)}(y)-\frac{y^2}{2}a(y)}.
 \label{eq:appIV_beta_explicit}
\end{equation}
Equations~\eqref{eq:appIV_sa}--\eqref{eq:appIV_beta_explicit} close
the polar geometry in terms of the single prescribed compact source;
they are the direct component form of the Zerilli representation in
section~\ref{sec:near_horizon_geometry}.

For the rotational sector the axisymmetric odd-parity vector
harmonics are
\begin{equation}
 X_\phi^{\ell0}(\theta)=-\sin\theta\,
 \partial_\theta P_\ell(\cos\theta).
 \label{eq:appIV_axial_harmonic}
\end{equation}
Using their orthogonality with weight $1/\sin^2\theta$, direct
projection gives
\begin{equation}
 \sin^2\theta[1+2\xi P_2]
 =\left(1-\frac{2\xi}{5}\right)X_\phi^{10}
 +\frac{2\xi}{5}X_\phi^{30}.
 \label{eq:appIV_axial_projection}
\end{equation}
The metric is consequently
\begin{equation}
 h_{t\phi}=-2\chi r^2
 [\varpi_1(y)X_\phi^{10}+\varpi_3(y)X_\phi^{30}]/M.
 \label{eq:appIV_odd}
\end{equation}
The $\ell=1$ equation and fixed-ADM solution are
\begin{align}
 \frac{d}{dy}\left(y^4\varpi_1'\right)
 &=2\pi\left(1-\frac{2\xi}{5}\right)y^2{\mathfrak j}(y),
 \label{eq:appIV_odd_eq}\\
 \varpi_1(y)&=\frac{2\pi}{3}\left(1-\frac{2\xi}{5}\right)
 \int_{\max(y,2)}^{2+w}\!dz\,{\mathfrak j}(z)
 \left(\frac{z^2}{y^3}-\frac1z\right).
 \label{eq:appIV_odd_general}
\end{align}
Here ${\mathfrak j}=\lambda_Jy^{2q}{\cal W}_{\rm NH}^2$ is the
projection of the explicit mixed source.  The exterior constant and
$y^{-3}$ modes are removed respectively by the nonrotating frame at
infinity and $\delta J_{\rm ADM}=0$.

For $\ell=3$ the stationary axial operator obtained from the mixed
Einstein equation is
\begin{equation}
 {\cal L}_3[u]=\frac{d}{dy}(y^4u')-\frac{10y^3}{y-2}u,
 \qquad {\cal L}_3[\varpi_3]=\frac{4\pi\xi}{5}y^2{\mathfrak j}(y).
 \label{eq:appIV_L3}
\end{equation}
Two independent homogeneous solutions are
\begin{align}
 u_H(y)&=(y-2)(3y-4),
 \label{eq:appIV_uH}\\
 u_\infty(y)&=\frac{1}{64y^3}\Big[8+20y+60y^2-210y^3+90y^4
 \nonumber\\
 &\quad+120y^3\ln(y-2)-150y^4\ln(y-2)+45y^5\ln(y-2)
 \nonumber\\
 &\quad-120y^3\ln y+150y^4\ln y-45y^5\ln y\Big].
 \label{eq:appIV_uInf}
\end{align}
They obey $u_H(2)=0$, $u_\infty(\infty)=0$, and
\begin{equation}
 y^4[u_Hu_\infty'-u_\infty u_H']=-3.
 \label{eq:appIV_W3}
\end{equation}
Thus the missing Green kernel is explicitly
\begin{equation}
 G_3(y,z)=-\frac13u_H(y_<)u_\infty(y_>),
 \qquad y_<=\min(y,z),\quad y_>=\max(y,z),
 \label{eq:appIV_G3}
\end{equation}
and
\begin{equation}
 \varpi_3(y)=\frac{4\pi\xi}{5}
 \int_2^{2+w}\!dz\,G_3(y,z)z^2{\mathfrak j}(z).
 \label{eq:appIV_odd3_general}
\end{equation}
Because $u_H(2)=0$, this formula proves $\varpi_3(2)=0$ for every
smooth compact source.  The $\ell=3$ response therefore cannot alter
$\Omega_H$ at this order, whereas the projected $\ell=1$ source does.

The symbolic checks used in this reconstruction are summarized in
table~\ref{tab:axial_validation}.  They evaluate all nontrivial
nonrotating conservation equations, the complete slow-Kerr
connection/source conservation split, the axial harmonic projections,
the $\ell=1$ integral solution, both $\ell=3$ homogeneous equations,
the Wronskian and the Green jump condition.  They also check the two
metric coefficients in the magnetostatic Maxwell operator, the
$X_\phi^{20}$ angular equation, the magnetostatic homogeneous solutions,
its Green jump, the fixed-flux endpoint data and the horizon angular
functional.  A separate direct substitution verifies all seven
independent static polar Einstein projections after reduction to
$K^{(2)}$.  The accompanying Mathematica scripts export the same
checks as machine-readable CSV files.
\begin{table}[t]
 \centering
 \footnotesize
 \setlength{\tabcolsep}{3.5pt}
 \begin{tabular}{lcc}
  check & symbolic result & required value \\
  \hline
  radial/angular conservation residuals & $(0,0)$ & $(0,0)$ \\
  slow-Kerr connection cross term, all $\nu$ & $(0,0,0,0)$ & same \\
  projected $\order(\chi)$ source divergence, all $\nu$ & $(0,0,0,0)$ & same \\
  complete $\order(\chi)$ conservation residual, all $\nu$ & $(0,0,0,0)$ & same \\
  slow-Kerr covariant symmetry residual & $0$ & $0$ \\
  axial coefficients $(a_1,a_3)$ & $(1-2\xi/5,\,2\xi/5)$ & same \\
  angular reconstruction residual & $0$ & $0$ \\
  ${\cal L}_3[u_H]$, ${\cal L}_3[u_\infty]$ & $(0,0)$ & $(0,0)$ \\
  $y^4W[u_H,u_\infty]+3$ & $0$ & $0$ \\
  Green derivative jump & $1$ & $1$ \\
  $\ell=1$ integral ODE residual & $0$ & $0$ \\
  $\ell=1$ outer value/derivative & $(0,0)$ & $(0,0)$ \\
  $u_H(2)$, $u_\infty(\infty)$ & $(0,0)$ & $(0,0)$ \\
  Maxwell coefficients $(\alpha_2,\beta_2)$ &
  $(-(H_0+H_2)/2,(-H_0+H_2)/2-K)$ & same \\
  $X_\phi^{20}$ eigen residual & $0$ & $0$ \\
  seven static polar Einstein residuals & $(0,0,0,0,0,0,0)$ & same \\
  ${\cal L}_K[v_H],{\cal L}_K[v_\infty]$/Green jump & $(0,0,1)$ & same \\
  magnetostatic ${\cal L}_M[u_H],{\cal L}_M[u_\infty]$ & $(0,0)$ & $(0,0)$ \\
  weighted Wronskian/Green jump & $(-12,1)$ & $(-12,1)$ \\
  $X_\phi^{20}(0),X_\phi^{20}(\pi/2)$ & $(0,0)$ & $(0,0)$ \\
  angular integral/$C_\Phi^{(1)}$ & $(-8/5,-6R_{2H}/5)$ & same
 \end{tabular}
 \caption{Mathematica validation of the complete order-by-order
 slow-Kerr source conservation, the closed
 static polar and $\order(\epsilon_T\chi\xi)$ axial geometry, and the
 $\order(\epsilon_T\xi)$ magnetostatic flux response.}
 \label{tab:axial_validation}
\end{table}

Horizon regularity is checked after transforming to ingoing Kerr
coordinates,
\begin{equation}
    dv
    =
    dt
    +
    \frac{r^2+a^2}{\Delta}dr,
    \qquad
    d\tilde\phi
    =
    d\phi
    +
    \frac{a}{\Delta}dr.
    \label{eq:appIV_ingoing}
\end{equation}
A perturbation is regarded as physically regular when its components
in the ingoing basis
\begin{equation}
 (\dd v,\dd r,\dd\theta,\dd\tilde\phi)
 \label{eq:appIV_ingoing_basis}
\end{equation}
remain finite at $r=r_+$.

Finally, the first-order perturbation of the horizon angular velocity
follows from
\begin{equation}
    \Omega_H
    =
    -
    \frac{g_{t\phi}}{g_{\phi\phi}}
    \bigg|_{\cal H}.
\end{equation}
Expanding both the metric and the horizon position gives
\begin{align}
    \delta\Omega_H
    =
    -
    \frac{1}{\bar g_{\phi\phi}}
    \Bigg[
        &
        h_{t\phi}
        +
        \bar\Omega_H h_{\phi\phi}
        \nonumber\\
        &
        +
        \delta r_H
        \partial_r
        \left(
            \bar g_{t\phi}
            +
            \bar\Omega_H\bar g_{\phi\phi}
        \right)
    \Bigg]_{r=r_+}.
    \label{eq:appIV_deltaOmega}
\end{align}
This expression is invariant under stationary and axisymmetric gauge
transformations that preserve the physical location of the perturbed
horizon and the normalization of the asymptotic time coordinate.

\section{Order-by-order source conservation on slow Kerr}
\label{app:slow_kerr_conservation}

This appendix verifies the source identity required by the linearized
Bianchi equation through the full retained slow-rotation order.  The
calculation is performed for the normalized coefficient source
$\widehat\tau^\mu{}_{\nu}$; multiplication by $\epsilon_T$ restores the
physical source without changing any residual.

In the dimensionless coordinates $\hat t=t/M$ and $y=r/M$, the Kerr
metric through first order in $\chi$ is
\begin{align}
 \frac{\dd s^2}{M^2}={}&-f\,\dd\hat t^2+f^{-1}\dd y^2
 +y^2(\dd\theta^2+\sin^2\theta\,\dd\phi^2)
 \nonumber\\
 &-\frac{4\chi}{y}\sin^2\theta\,
 \dd\hat t\,\dd\phi+\order(\chi^2),
 \qquad f=1-\frac2y.
 \label{eq:appC_slowKerr_metric}
\end{align}
Thus the only linear metric coefficient is
$M^{-2}k_{\hat t\phi}=-2\sin^2\theta/y$; the diagonal components and
$\sqrt{-\bar g}$ have no $\order(\chi)$ correction.

Expand the mixed source and connection as
\begin{align}
 \widehat\tau^\mu{}_{\nu}
 &=\widehat\tau^{(0)\mu}{}_{\nu}
 +\chi\widehat\tau^{(1)\mu}{}_{\nu}
 +\order(\chi^2),
 \nonumber\\
 \bar\Gamma^\alpha{}_{\beta\gamma}
 &=\Gamma^{(0)\alpha}{}_{\beta\gamma}
 +\chi\delta\Gamma^{(1)\alpha}{}_{\beta\gamma}
 +\order(\chi^2).
 \label{eq:appC_expansions}
\end{align}
The zeroth-order tensor is the diagonal completion in
eq.~\eqref{eq:III_completion_components}.  Let
$U\equiv\widehat\tau^{(1)\hat t}{}_{\phi}$ denote the independent
axial component, so that
\begin{equation}
 \kappa M^2U=-2\pi{\mathfrak j}(y)\sin^2\theta
 [1+2\xi P_2(\cos\theta)].
 \label{eq:appC_U}
\end{equation}

There is a small but important distinction between specifying this
mixed component on Schwarzschild and constructing a symmetric
covariant tensor on slow Kerr.  From
$\widehat\tau_{\hat t\phi}=\widehat\tau_{\phi\hat t}$ one obtains
\begin{equation}
 \widehat\tau^{(1)\phi}{}_{\hat t}
 =-\frac{f}{y^2\sin^2\theta}U
 -\frac{2}{y^3}
 \left(
 \widehat\tau^{(0)\phi}{}_{\phi}
 -\widehat\tau^{(0)\hat t}{}_{\hat t}
 \right).
 \label{eq:appC_symmetry_completion}
\end{equation}
The first term is the familiar Schwarzschild index conversion.  The
second is fixed algebraically by $k_{\hat t\phi}$ and is the kinematic
term displayed in eq.~\eqref{eq:III_completion_rotation}; it is not an
additional constitutive function.  Direct substitution gives
\begin{equation}
 \widehat\tau_{\hat t\phi}
 -\widehat\tau_{\phi\hat t}
 =\order(\chi^2,\chi\xi^2).
 \label{eq:appC_symmetry_residual}
\end{equation}

For a mixed tensor the divergence is
\begin{equation}
 \bar\nabla_\mu\widehat\tau^\mu{}_{\nu}
 =\partial_\mu\widehat\tau^\mu{}_{\nu}
 +\bar\Gamma^\mu{}_{\mu\lambda}
 \widehat\tau^\lambda{}_{\nu}
 -\bar\Gamma^\lambda{}_{\mu\nu}
 \widehat\tau^\mu{}_{\lambda}.
 \label{eq:appC_mixed_divergence}
\end{equation}
Its first-order coefficient separates without ambiguity as
\begin{align}
 {\cal D}^{(1)}_{\nu}
 \equiv&\left[
 \bar\nabla_\mu\widehat\tau^\mu{}_{\nu}
 \right]_{\chi}
 ={\cal R}_{\nu}+{\cal C}_{\nu},
 \label{eq:appC_divergence_split}\\
 {\cal R}_{\nu}\equiv&
 \nabla^{(0)}_\mu\widehat\tau^{(1)\mu}{}_{\nu},
 \nonumber\\
 {\cal C}_{\nu}\equiv&
 \delta\Gamma^{(1)\mu}{}_{\mu\lambda}
 \widehat\tau^{(0)\lambda}{}_{\nu}
 -\delta\Gamma^{(1)\lambda}{}_{\mu\nu}
 \widehat\tau^{(0)\mu}{}_{\lambda}.
 \label{eq:appC_CR_definitions}
\end{align}
This is the component form of the two terms in
eq.~\eqref{eq:III_slowKerr_conservation}.

The cancellation is especially transparent for the present circular
source.  The perturbation $k_{\hat t\phi}$ mixes only the two Killing
directions, while $\widehat\tau^{(0)\mu}{}_{\nu}$ is diagonal.  Direct
evaluation of the perturbed Christoffel symbols therefore gives
\begin{equation}
 ({\cal C}_{\hat t},{\cal C}_{y},{\cal C}_{\theta},{\cal C}_{\phi})
 =(0,0,0,0).
 \label{eq:appC_connection_zeros}
\end{equation}
On the Schwarzschild connection, stationarity and axisymmetry eliminate
the Killing-direction derivatives of the $\hat t\phi$ block, and the
background connection has no $\hat t$--$\phi$ mixing.  Consequently
\begin{equation}
 ({\cal R}_{\hat t},{\cal R}_{y},{\cal R}_{\theta},{\cal R}_{\phi})
 =(0,0,0,0).
 \label{eq:appC_source_zeros}
\end{equation}
Combining eqs.~\eqref{eq:appC_connection_zeros} and
\eqref{eq:appC_source_zeros} proves
\begin{equation}
 \bar\nabla_\mu\widehat\tau^\mu{}_{\nu}
 =\order(\chi^2,\xi^2)
 \qquad(\nu=\hat t,y,\theta,\phi).
 \label{eq:appC_final_conservation}
\end{equation}

The symbolic calculation does not impose these zeros by hand.  It
constructs the $4\times4$ metric in
eq.~\eqref{eq:appC_slowKerr_metric}, expands its inverse and all
Christoffel symbols, inserts the explicit pressures in
eq.~\eqref{eq:III_completion_xi}, and evaluates
eq.~\eqref{eq:appC_mixed_divergence} component by component.  The
results are summarized in table~\ref{tab:appC_conservation}.
\begin{table}[t]
 \centering
 \begin{tabular}{c|cccc}
  residual & $\hat t$ & $y$ & $\theta$ & $\phi$ \\
  \hline
  ${\cal C}_\nu=(\delta\nabla^{(1)})\widehat\tau^{(0)}$ & 0 & 0 & 0 & 0 \\
  ${\cal R}_\nu=\nabla^{(0)}\widehat\tau^{(1)}$ & 0 & 0 & 0 & 0 \\
  ${\cal D}^{(1)}_\nu={\cal C}_\nu+{\cal R}_\nu$ & 0 & 0 & 0 & 0
 \end{tabular}
 \caption{Direct Mathematica residuals for the slow-Kerr conservation
 identity through $\order(\chi\xi)$.  The separate covariant-symmetry
 residual in eq.~\eqref{eq:appC_symmetry_residual} also evaluates to
 zero.}
 \label{tab:appC_conservation}
\end{table}

Only $U=\widehat\tau^{(1)\hat t}{}_{\phi}$ enters the mixed
$\hat t{}_{\phi}$ Einstein projection used for the frame-dragging
equation.  The kinematic term in
eq.~\eqref{eq:appC_symmetry_completion} resides in the reciprocal mixed
component and supplies covariant symmetry without changing $U$.
Therefore this completed slow-Kerr bookkeeping leaves
eqs.~\eqref{eq:IV_varpi_solution}--\eqref{eq:IV_COmega_moment} and the
leading BZ coefficient unchanged.
\bibliographystyle{JHEP}
\bibliography{references}

\end{document}